\documentclass[a4paper, fleqn, usenatbib, useAMS]{mnras}
\usepackage[T1]{fontenc}
\usepackage{ae, aecompl}			
\usepackage{newtxtext, newtxmath} 	
\usepackage{mathtools}
\usepackage{graphicx}
\usepackage{amsmath}
\usepackage{amssymb}
\usepackage{multicol}
\usepackage{bm}
\usepackage{pdflscape}
\usepackage{natbib}
\usepackage[section]{placeins}
\usepackage{lipsum}
\usepackage{etoolbox} 
\usepackage{tabularx} 
\usepackage{xcolor}
\hypersetup{
	colorlinks 		= true, 
	urlcolor		= blue, 
	linkcolor		= blue, 
	citecolor 		= blue
}

\newcommand{\ddfrac}[2]{\frac{\displaystyle #1}{\displaystyle #2}} 
\newcommand{\refp}[1]{(\ref{#1})} 
\graphicspath{ {../plots/} }

\title[Impact of Starbursts on Abundances]{The Impact of Starbursts on Element 
Abundance Ratios}

\author[J.W. Johnson \& D.H. Weinberg]{
	James W. Johnson\thanks{Contact e-mail: \href{mailto:
	johnson.7419@osu.edu}{johnson.7419@osu.edu}} and 
	David H. Weinberg
	\\
	Department of Astronomy, The Ohio State University, 140 W. 18th Ave. 
	Columbus, OH 43210, USA
}

\date{Accepted XXX; Received YYY; in original form ZZZ}
\pubyear{2018}

\begin{document}
\label{firstpage}
\pagerange{\pageref{firstpage}--\pageref{lastpage}}
\maketitle

\defcitealias{Weinberg2017}{WAF17} 
\defcitealias{Andrews2017}{AWSJ17} 

\begin{abstract}
We investigate the impact of bursts in star formation on the predictions of 
one-zone chemical evolution models, adopting oxygen (O), iron (Fe), and 
strontium (Sr), as representative $\alpha$, iron-peak, and s-process 
elements, respectively. To this end, we develop and make use of the 
\texttt{Versatile Integrator for Chemical Evolution} (\texttt{VICE}), a 
\texttt{python} package designed to allow flexible descriptions of star 
formation histories, accretion, and outflow histories and nucleosynthetic 
yields. Starbursts driven by a temporary boost of gas accretion rate create 
loops in [O/Fe]-[Fe/H] evolutionary tracks and a peak in the stellar [O/Fe] 
distribution at intermediate values. Bursts driven by a temporary boost of 
star formation efficiency have a similar effect, and they also produce a 
population of $\alpha$-deficient stars during the depressed star formation 
phase that follows the burst. This $\alpha$-deficient population is more 
prominent if the outflow rate is tied to a time-averaged star formation rate 
(SFR) instead of the instantaneous SFR. Theoretical models of Sr production 
predict a strong metallicity dependence of supernova and asymptotic giant 
branch (AGB) star yields, though comparison to data suggests an additional 
source that is nearly metallicity-independent. Evolution of [Sr/Fe] and [Sr/O] 
during a starburst is complex because of the yield metallicity dependence 
and the multiple timescales in play. Moderate amplitude (10-20\%) sinusoidal 
oscillations in SFR produce loops in [O/Fe]-[Fe/H] tracks and multiple peaks in 
[O/Fe] distributions, which could be one source of intrinsic scatter in 
observed sequences. We investigate models that have a factor of~$\sim$2 
enhancement of SFR at $t$ = 12 Gyr, as suggest by some recent Milky Way 
observations. A late episode of enhanced star formation could help explain 
the existence of young stars with moderate $\alpha$-enhancements and the 
surprisingly young median age found for solar metallicity stars in the 
solar neighborhood, while also raising the possibility that this starburst has 
not yet fully decayed. \texttt{VICE} is publicly available at 
<\url{http://github.com/giganano/VICE.git}>. 
\end{abstract}


\section{Introduction}
\label{sec:intro}
The elemental abundances and abundance ratios of stars encode information 
about the history of galactic enrichment and about the stellar processes that 
produce the elements. The ratio of $\alpha$-element abundances to the iron 
abundance is an especially important diagnostic, because the $\alpha$-elements 
(e.g. O, Mg, and Si) are produced primarily by massive stars with short 
lifetimes, while Fe is also produced in substantial amounts by Type Ia 
supernovae (SNe Ia) that explode after a wide range of delay times. In simple 
chemical evolution models with smooth star formation histories, the track of 
[$\alpha$/Fe] vs. [Fe/H]\footnote{
	We follow conventional notation in which [X/Y] $\equiv\ \log_{10}(X/Y) 
	- \log_{10}(X_\odot/Y_\odot)$.  
} first follows a plateau that reflects the IMF\footnote{
	IMF: Initial Mass Function 
}-averaged yield of core collapse supernovae (CCSNe), then turns downward as 
SNe Ia begin to add Fe without associated $\alpha$-elements. If the model has 
continuing gas accretion, then the [Fe/H] and [$\alpha$/Fe] ratios tend to 
approach an equilibrium in which the production of new elements is balanced by 
dilution from freshly accreted gas and by depletion of metals from new star 
formation or outflows (\citealp[][hereafter 
\citetalias{Andrews2017}]{Larson1972, Finlator2008, Andrews2017}; 
\citealp[][hereafter \citetalias{Weinberg2017}]{Weinberg2017}). 
\par\null\par\null\par 
In this paper we examine the impact of starbursts - sudden and temporary 
increases in the star formation rate - which perturb chemical evolution by 
temporarily boosting the rate of CCSNe relative to SNe Ia from earlier 
generations of stars. We adopt one-zone evolution models in which stars form 
from and enrich a fully mixed gas reservoir subject to accretion and outflow
(see, e.g. \citealp{Schmidt1959, Schmidt1963, Larson1972, Tinsley1980} for 
classical examples; \citetalias{Weinberg2017}, \citetalias{Andrews2017} for 
more recent work). Although idealized, one-zone models may be a reasonable 
approximation for the evolution of dwarf galaxies. The Milky Way can be 
modeled as an annular sequence of one-zone models ~\citep{Matteucci1989}, 
which may be connected by processes such as the radial migration of 
stars~\citep{Schoenrich2009, Minchev2017} and radial gas 
flows~\citep{Lacey1985, Bilitewski2012}. 
\par 
The [$\alpha$/Fe]-[Fe/H] tracks observed in the inner Milky Way agree well with 
the predictions of a one-zone model in which the star-forming gas disk 
contracts vertically over a period of~$\sim$2 Gyr~\citep{Hayden2015, 
Freudenburg2017}. In the solar neighborhood, stars with high and low vertical 
velocities trace distinct [$\alpha$/Fe]-[Fe/H] tracks known as the 
``high-$\alpha$'' and ``low-$\alpha$'' sequences~\citep{Bensby2003, 
Hayden2015}, a bimodality whose origin is still not fully understood. 
\citetalias{Andrews2017} and~\citetalias{Weinberg2017} systematically 
investigate the influence of model parameters on the [$\alpha$/Fe]-[Fe/H] 
tracks of one-zone models with smooth star formation histories, with 
particlular attention to the role of outflows in regulating the equilibrium 
metallicity. In agreement with previous studies of the galaxy mass-metallicity 
relation~\citep[e.g. ][]{Dalcanton2007, Finlator2008, Peeples2011, Zahid2012}, 
they find that achieving a solar metallicity interstellar medium (ISM) requires 
strong outflows, with a mass-loading factor $\eta\equiv\dot{M}_\text{out}/
\dot{M}_*\approx2.5$ for a~\citet{Kroupa2001} IMF where every star of mass 
8 - 100 $M_\odot$ explodes as a CCSN with the yields predicted 
by~\citet{Chieffi2004,Chieffi2013}. 
\par 
\citet{Gilmore1991} investigated the impact of a bursty star formation history 
on [O/Fe]-[Fe/H] tracks, focusing on application to the Large Magellanic 
Cloud. \citetalias{Weinberg2017} investigated a model in which a sudden change 
of star formation efficiency induces a starburst, causing an upward jump in 
[O/Fe] followed by a return to equilibrium (see their Figure 9). In this 
paper we study the impact of starbursts more systematically, showing the 
different forms of [O/Fe]-[Fe/H] tracks and stellar [O/Fe] distributions for 
bursts induced by a sudden influx of gas, a boost in gas accretion rate, or 
an increase of star formation efficiency. We also investigate the 
connection between starbursts and winds, considering the possibility that 
outflows are tied to a time-averaged star formation rate (SFR) instead of the 
instantaneous SFR that governs the rate of CCSN enrichment. In addition to 
O and Fe, we examine strontium (Sr) as a representative element that has both 
a CCSN contribution and an asymptotic giant branch (AGB) star contribution 
with a metallicity dependent yield. 
\par 
To this end we have developed a publicly available\footnote{
	\url{https://github.com/giganano/VICE.git}
} \texttt{python} package, the \texttt{Versatile Integrator for Chemical 
Evolution} (\texttt{VICE}), which solves the integro-differential equations of 
a one-zone chemical evolution model. 
Compared to \texttt{flexCE}~\citepalias{Andrews2017}\footnote{
	\url{https://github.com/bretthandrews/flexce}
}, \texttt{VICE} has a simpler methodology in that it works directly from 
user-specified IMF-averaged yields rather than drawing CCSNe stochastically 
from the IMF of massive stars. \texttt{VICE} focuses instead on versatility 
in specifying star formation histories, gas accretion histories, and star 
formation laws as arbitrary functions of time, and it will automatically 
compute yield tables from a variety of sources if requested~\citep[e.g.][among 
others to be added in subsequent versions]{Woosley1995, Iwamoto1999, 
Chieffi2004, Karakas2010, Cristallo2011, Seitenzahl2013, Limongi2018}. With a 
backend written in \texttt{C}, \texttt{VICE} also achieves powerful computing 
speeds while maintaining this level of flexibility. We anticipate adding 
further capabilities to \texttt{VICE} in the future, including extensions to 
multizone models. 
\par 
Our models in this paper are motivated primarily by considerations of dwarf 
galaxies, which often show evidence of bursty star formation 
histories~\citep[e.g.][]{Weisz2011, Weisz2014}. However, even local variations 
in star formation induced by passage of gas through a spiral arm can induce 
some of these effects, damped mainly by the fact that such events typically 
convert only a small fraction of the available gas into 
stars~\citepalias{Weinberg2017}. In their hydrodynamic simulations of disk 
galaxy formation,~\citet{Clarke2019} find that massive clumps in young 
gas-rich disks convert much of their gas into stars and therefore self-enrich, 
following tracks in [$\alpha$/Fe]-[Fe/H] space that resemble those of our 
efficiency-induced starburst models below. They propose that a superposition 
of such bursts is responsible for the high-$\alpha$ sequence observed in the 
Milky Way [$\alpha$/Fe]-[Fe/H] diagram. In addition to bursts, we investigate 
here the effect of slow sinusoidal variations in SFR, finding that these less 
dramatic variations could produce scatter in [$\alpha$/Fe] at fixed [Fe/H], at 
least along the low-$\alpha$ sequence. 

\section{Methods: The One-Zone Approximation}
\label{sec:methods}
\subsection{The Gas Supply, Star Formation Rate, and Star Formation Efficiency}
Under the one-zone approximation, the fundamental assumption is instantaneous 
mixing of newly released metals throughout the star-forming gas reservoir. 
This reduces the equations of GCE to a system of integro-differential 
equations of mass with time, which can be integrated numerically. Under this 
formalism the time-derivative of the gas-supply is given by: 
\begin{subequations}\begin{align} 
\label{eq:mdot_gas} 
\dot{M}_\text{g} &= \dot{M}_\text{in} - \dot{M}_* - \dot{M}_\text{out} + 
\dot{M}_\text{returned} \\ 
&\approx \dot{M}_\text{in} - \dot{M}_*(1 + \eta - r_\text{inst}) \\ 
&= \dot{M}_\text{in} - M_\text{g}\tau_*^{-1}(1 + \eta - r_\text{inst}) 
\end{align}\end{subequations} 
where $\dot{M}_\text{in}$ is the mass infall rate, $\dot{M}_*$ is the SFR, 
$\dot{M}_\text{out}$ is the mass outflow rate, and $\dot{M}_\text{returned}$ 
is the rate of recycling. The star formation efficiency (SFE) timescale is 
defined by $\tau_* = M_\text{g}/\dot{M}_*$, and the parameters $\eta$ and 
$r_\text{inst}$ are discussed further below. 
\texttt{VICE} allows the user to specify the initial gas supply $M_\text{g,0}$ 
and the inflow rate $\dot{M}_\text{in}$ as a function of time, in which case 
the SFR follows from the star formation law $\dot{M}_* = M_\text{g}/\tau_*$. 
Alternatively, the user can specify the star formation history $\dot{M}_*$ 
itself or the gas supply at all times $M_\text{g}(t)$, with the star formation 
law supplying the remaining quantity. In these cases, the infall rate is 
determined implicitly by solving equation~\refp{eq:mdot_gas}. The former 
approach is somewhat more common in chemical evolution modeling, reflecting 
the expectation that a galaxy's star formation history is ultimately governed 
by the rate at which it accretes gas from the surrounding circumgalactic 
medium. However, a galaxy's star formation history can be estimated 
observationally while its accretion history cannot, and for analytic solution 
it is often more convenient to specify $\dot{M}_*(t)$ rather than 
$\dot{M}_\text{in}(t)$ as shown by~\citetalias{Weinberg2017}. For the 
calculations in this paper, we specify $\dot{M}_\text{in}(t)$ and allow the 
SFR to follow from the gas supply unless otherwise specified. 
\par 
As a default value for the SFE timescale we adopt 
$\tau_*$ = 2 Gyr, the typical value found for molecular gas in a wide range of 
star-forming galaxies~\citep{Leroy2008}. The observationally inferred $\tau_*$ 
is lower in some starbursting systems, as short as~$\sim$100 Myr; however, the 
details of this relation are subject to the ongoing debate about the 
CO-to-H$_2$ conversion 
factor~\citep[for details, see the review in][]{Kennicutt2012}. 
Relative to the total gas supply, the SFE timescale will be longer if much of 
the reservoir is in atomic form, roughly 
$\tau_* = (\text{2 Gyr})(1 + M_\text{HI}/M_{\text{H}_2})$. 
\texttt{VICE} allows the 
user to specify $\tau_*$ as a function of time, simultaneously allowing it to 
vary with the gas supply according to the Kennicutt-Schmidt 
relation~\citep{Schmidt1959, Schmidt1963, Kennicutt1998}. If one views the 
gas reservoir as representing an annulus of a disk, with gas surface density 
$\Sigma_\text{g} = M_\text{g}/A_\text{ann}$, then the classic non-linear 
Kennicutt-Schmidt law $\dot{\Sigma}_* \propto \Sigma_\text{g}^{1.5}$ implies 
$\tau_* \propto M_\text{g}^{-0.5}$. We adopt this form in some of our 
calculations below. 

\begin{table*} 
\caption{Galactic chemical evolution parameters and their fiducial/unperturbed 
values adopted in this paper (if applicable). For further details 
on each parameter, see \texttt{VICE}'s science documentation, available 
at~\url{https://github.com/giganano/VICE/tree/master/docs}. } 
\begin{tabularx}{\textwidth}{l @{\extracolsep{\fill}} l r} 
\hline 
\hline 
Quantity & Description & Fiducial/Unperturbed Value \\ 
\hline 
$M_\text{g}$ & Gas Supply & $\sim6\times10^9\ M_\odot$ \\ 
$\dot{M}_*$ & Star Formation Rate & $\sim3\ M_\odot\ \text{yr}^{-1}$ \\ 
$\dot{M}_\text{in}$ & Infall Rate & $\sim9\ M_\odot\ \text{yr}^{-1}$ \\ 
$\dot{M}_\text{out}$ & Outflow Rate & 
	$\eta\langle\dot{M}_*\rangle_{\tau_\text{s}}$ \\ 
$\dot{M}_\text{returned}$ & Recycling Rate & 
	Continuous (see equation~\refp{eq:mdot_returned}) \\ 
$\tau_*$ & Star Formation Efficiency (SFE) Timescale ($M_\text{g}/\dot{M}_*$) & 
	2 Gyr \\ 
$\eta$ & Mass-Loading Factor ($\dot{M}_\text{out} / \dot{M}_*$) & 2.5 \\ 
$\xi_\text{enh}$ & Outflow Enhancement Factor ($Z_\text{out}/Z_\text{ISM}$) & 
	1 \\ 
$\dot{M}_x^\text{CC}$ & Rate of Enrichment from CCSNe & N/A \\ 
$y_x^\text{CC}$ & IMF-integrated fractional yield from CCSNe & 
	O: 0.015; Fe: 0.0012; Sr: $3.5\times10^{-8}$ \\ 
$\dot{M}_x^\text{Ia}$ & Rate of Enrichment from SNe Ia & N/A \\ 
$y_x^\text{Ia}$ & IMF-integrated fractional yield from SNe Ia & 
	O: 0.0; Fe: 0.0017; Sr: 0.0 \\ 
$\dot{M}_x^\text{AGB}$ & Rate of Enrichment from AGB stars & N/A \\ 
$y_x^\text{AGB}(m_\text{to} | Z)$ & Fraction yield from an AGB star of mass 
$m_\text{to}$ and metallicity $Z$ & \citet{Cristallo2011} \\ 
$r(t)$ & Cumulative Return Fraction (CRF) & N/A \\ 
$h(t)$ & Main Sequence Mass Fraction (MSMF) & N/A \\ 
$Z_\text{ISM}$ & Total Metallicity by Mass of the ISM & N/A \\ 
\hline
\end{tabularx}
\label{tab:docs} 
\end{table*} 

\subsection{The Cumulative Return Fraction} 
\label{sec:crf} 
The cumulative return fraction (CRF) $r(t)$ is the fraction of a stellar 
population's mass formed at $t$ = 0 that has been returned to the ISM by a 
time $t$ through stellar winds or supernova explosions. In \texttt{VICE}, we 
calculate $r(t)$ approximately by assuming that all stars with initial mass 
$M > 8 M_\odot$ leave a remnant of 1.44 $M_\odot$ while those less than 8 
$M_\odot$ leave remnants of mass $0.394 M_\odot + 0.109 M$~\citep{Kalirai2008}. 
In these calculations, the main sequence turnoff mass at a time $t$ following 
the formation of a stellar population is assumed to be $M_\text{to}/M_\odot 
\approx (t/\text{10 Gyr})^{-1/3.5}$, the same form as adopted 
in~\citetalias{Weinberg2017}. While this formula is less accurate for high 
$M_\text{to}$, the return timescale for these stars is much shorter than 
other chemical evolution timescales anyway, so the approximation is adequate. 
\par
\texttt{VICE} calculates the time-dependent return rate from all previous 
stellar generations as: 
\begin{equation}
\label{eq:mdot_returned} 
\dot{M}_\text{returned}(t) = \int_0^t\dot{M}_*(t - t')\dot{r}(t')dt' . 
\end{equation} 
Alternatively, one can make the approximation that all mass (from AGB stars as 
well as CCSNe) is returned instantaneously, in which case: 
\begin{equation} 
\label{eq:mdot_returned_inst}
\dot{M}_\text{returned} = r_\text{inst}\dot{M}_* . 
\end{equation}
For a Kroupa IMF, the CRF is $r(t) \approx$ 0.37, 0.40, and 0.45 after 1, 2, 
and 10 Gyr, and~\citetalias{Weinberg2017} shows that the difference between 
[$\alpha$/Fe]-[Fe/H] evolution with the time-dependent return of 
equation~\refp{eq:mdot_returned} and the instantaneous approximation with 
$r_\text{inst}$ = 0.4 is very small. Nonetheless, numerical implementation of 
equation~\refp{eq:mdot_returned} is neither difficult nor time-consuming, so we 
use continuous recycling throughout this paper. We note that 
equation~\refp{eq:mdot_returned_inst} is not equivalent to the ``instantaneous 
recycling approximation'' as that term is most frequently used, where it 
implies instantaneous return of \textit{newly produced} elements as well as the 
mass and metals that stars are born with. The full instantaneous recycling 
approximation is accurate for pure-CCSN elements if the star formation 
history is smooth on timescales of~$\sim$100 Myr, but it is not an accurate 
description of SN Ia enrichment. 

\subsection{The Mass Loading Factor} 
For the outflow mass loading factor $\eta$ we adopt a default value of 2.5, 
the same as~\citetalias{Weinberg2017}, with the result that our models 
approach approximately solar abundances at late times given our adopted CCSN 
and SN Ia yields. However, as noted in \S~\ref{sec:intro}, we also 
consider the possibility that the outflow rate is not tied to the 
instantaneous SFR but to some time-averaged value. This introduces an 
additional parameter, the smoothing timescale $\tau_\text{s}$, defined such 
that 
\begin{equation}\begin{split}
\label{eq:tau_s}
\dot{M}_\text{out} &= \eta\langle\dot{M}_*\rangle_{\tau_\text{s}} \\ 
&= \begin{dcases}
\frac{\eta}{\tau_\text{s}}\int_{t - \tau_\text{s}}^{t} \dot{M}_*(t') dt' & 
(t > \tau_\text{s}) \\ 
\frac{\eta}{t}\int_{0}^{t} \dot{M}_*(t') dt' & (0 \leq t \leq \tau_\text{s}) . 
\end{dcases}
\end{split}\end{equation} 
If galactic winds are driven primarily by massive star winds, radiation 
pressure, and CCSNe, then the effective smoothing timescale is likely to be 
short ($\tau_\text{s}\sim$ 50 Myr), and smoothing will have little impact on 
chemical evolution if the SFR is smooth on these timescales. However, if SNe 
Ia play a central role in driving winds, then effective smoothing times as 
long as $\tau_\text{s}\sim$ 1 Gyr are possible, altering the relative ejection 
of CCSNe and SNe Ia elements from a shorter duration starburst. Cosmic ray 
feedback could also produce an intermediate smoothing time, because the 
energy deposited by CCSNe can be temporarily stored in cosmic rays before 
building up sufficiently to drive a wind. While \texttt{VICE} allows the user 
to specify $\eta$ as a function of time, we do not consider models with a 
time-varying $\eta$ here. 

\subsection{CCSNe} 
Following~\citetalias{Weinberg2017}, we implement in \texttt{VICE} the 
instantaneous explosion approximation to CCSNe. This is a good approximation, 
because the lifetimes of CCSN progenitors ($\lesssim$40 Myr for the least 
massive ones) are much shorter than the relevant timescales of galactic 
chemical evolution (GCE). In our models, a given yield of some element X is 
ejected simultaneously with the formation of a stellar population at all 
timesteps, thus implying the mathematical form: 
\begin{equation} 
\label{eq:mdot_ccsne}
\dot{M}_\text{x}^\text{CC} = y_\text{x}^\text{CC}(Z)\dot{M}_* 
\end{equation} 
where $y_\text{x}^\text{CC}$ is the fraction of the stellar population's 
\textit{total} mass that is converted to the element $x$ at a metallicity 
$Z$. \texttt{VICE} allows any numerical value as well as user-constructed 
functions of $Z$ to describe $y_\text{x}^\text{CC}$ for all elements. 
We discuss our adopted CCSN yields in \S~\ref{sec:yields}. 

\subsection{The SN Ia Delay-Time Distribution (DTD)} 
\label{sec:dtd} 
We define $R_\text{Ia}(t)$ to be the rate of SNe Ia per unit stellar mass 
formed at a time $t$ following an episode of star formation. 
Following~\citetalias{Weinberg2017} (see Appendix A therein), we set: 
\begin{equation} 
\label{eq:mdot_ia} 
M_\text{x}^\text{Ia} = y_\text{x}^\text{Ia}\langle\dot{M}_*\rangle_\text{Ia} 
\end{equation} 
where 
\begin{equation} 
\label{eq:y_x_Ia} 
y_\text{x}^\text{Ia} \equiv m_\text{x}^\text{Ia} 
\int_{t_\text{D}}^{t_\text{max}} R_\text{Ia}(t)dt = 
m_\text{x}^\text{Ia}\frac{N_\text{Ia}}{M_*} 
\end{equation} 
is the fractional yield of element X from all SNe Ia that would explode 
between the minimum delay time $t_\text{D}$ and a specified maximum time 
$t_\text{max}$. Here $m_\text{x}^\text{Ia}$ is the average mass yield of the 
element X per SN Ia, $M_*$ is the mass of the stellar population, and 
\begin{equation}
\label{eq:mdotstarIa} 
\langle\dot{M}_*\rangle_\text{Ia} \equiv \ddfrac{
	\int_0^t \dot{M}_*(t - t')R_\text{Ia}(t')dt' 
}{
	\int_{t_\text{D}}^{t_\text{max}} R_\text{Ia}(t')dt' 
} 
\end{equation} 
is the time-averaged SFR weighted by the SNe Ia DTD. In implementation, 
\texttt{VICE} enforces $t_\text{max}$ = 15 Gyr always, though provided one is 
consistent in equations~\refp{eq:y_x_Ia} and~\refp{eq:mdotstarIa}, the result 
of equation~\refp{eq:mdot_ia} is independent of the choice of $t_\text{max}$. 
This formulation implicitly assumes that $R_\text{Ia}$ and 
$m_\text{x}^\text{Ia}$ are independent of the birth population's metallicity. 
As discussed further in \S~\ref{sec:yields} below, we adopt a power-law DTD 
with $R_\text{Ia} \propto t^{-1.1}$ and a minimum time delay of 
$t_\text{D}$ = 150 Myr. \texttt{VICE} allows the user to specify alternative 
forms for the DTD, including user-constructed functional forms. 

\subsection{AGB Stars} 
For AGB enrichment, we implement in \texttt{VICE} an algorithm that tracks the 
mass rate of change of a single stellar population to determine the mass in 
dying stars at each timestep. The rate of mass enrichment of an element $x$ 
from AGB stars is then given by 
\begin{equation} 
\label{eq:mdot_agb} 
\dot{M}_\text{x}^\text{AGB} = -\int_0^t y_\text{x}^\text{AGB}(m_\text{to}
(t - t'), Z_\text{ISM}(t'))\dot{M}_*(t')\dot{h}(t - t')dt' 
\end{equation} 
where $y_\text{x}^\text{AGB}$ is the yield of a star of mass $m$ and 
total metallicity $Z$, and $h$ is the main sequence mass fraction, 
defined to be the fraction of a stellar population's mass that is in the form 
of main sequence stars at a time $t$ following its formation. By definition, 
$h$ = 1 at $t$ = 0, and declines monotonically; hence the minus sign in 
equation~\refp{eq:mdot_agb}. $h$ is fully described by the adopted 
stellar IMF and the mass-lifetime relation (see \texttt{VICE}'s Science 
Documentation for further details). 

\subsection{Adopted Nucleosynthetic Yields} 
\label{sec:yields} 
For CCSN yields of O and Fe, we adopt the same values 
as~\citetalias{Weinberg2017}, $y_\text{O}^\text{CC}$ = 0.015 and 
$y_\text{Fe}^\text{CC}$ = 0.0012, independent of metallicity. The former is 
approximately the value computed from the yields of~\citet{Chieffi2004} 
for solar metallicity stars assuming a Kroupa IMF in which all stars with M = 
8 - 100 $M_\odot$ explode. CCSN iron yields are difficult to predict from 
first principles; our choice yields a plateau at [O/Fe] $\approx$ +0.45, in 
reasonable agreement with observations. Although we investigate Sr as a 
representative example of an AGB element, it is also expected to have a CCSN 
contribution. In \S~\ref{sec:sr} we examine the impact of various assumptions 
of the form of $y_\text{Sr}^\text{CC}$, including one with no 
metallicity dependence, one that depends linearly on $Z$, another with a 
$y_\text{Sr}^\text{CC} \propto 1 - e^{-kZ}$ dependence, and one in which 
$y_\text{Sr}^\text{CC} = 0$ as a limiting case describing pure AGB enrichment. 
\par
For the SN Ia iron yield we adopt $y_\text{Fe}^\text{Ia}$ = 0.0017, similar to 
the values used by~\citet{Schoenrich2009},~\citetalias{Andrews2017}, 
and~\citetalias{Weinberg2017}. This value is based on a normalization of the 
SNe Ia DTD that yields $N_\text{Ia}/M_* = \int_{t_\text{D}}^{t_\text{max}}
R_\text{Ia}(t)dt = 2.2\times10^{-3} M_\odot^{-1}$, consistent with 
$(2\pm1)\times10^{-3} M_\odot^{-1}$ from~\citet{Maoz2012}, and 
$m_\text{Fe}^\text{Ia}$ = 0.78 $M_\odot$ from the W70 explosion model 
of~\citet{Iwamoto1999}. Because this enrichment channel is negligible for 
O and Sr, we adopt $y_\text{Sr}^\text{Ia}$ and $y_\text{O}^\text{Ia}$ = 0 
throughout this work. As noted in \S~\ref{sec:dtd}, we adopt a 
$t^{-1.1}$ power-law DTD, again motivated by~\citet{Maoz2012}, with a minimum 
delay time of $t_\text{D}$ = 150 Myr. In principle, $t_\text{D}$ could be as 
short as the lifetime of the most massive stars that produce white dwarfs 
(roughly 40 Myr), but it is not clear empirically whether the t$^{-1.1}$ 
power-law extends to such small $t$. As a rule of thumb, it is useful to 
remember that a $t^{-1}$ power-law DTD would yield equal numbers of SNe Ia 
per logarithmic time interval (i.e. the same number between 0.1 - 1 Gyr and 
1 - 10 Gyr). Thus 1 Gyr is the approximate characteristic time for half of the 
SN Ia iron to be produced. If $t_\text{D}$ is as short as 0.05 Gyr, then 
about 20\% of SNe Ia explode between 0.05 and 0.15 Gyr, enough to noticably 
shift the ``knee'' of the [$\alpha$/Fe]-[Fe/H] tracks. For our default 
$t_\text{D}$ = 0.15 Gyr, these tracks are nearly identical to those of an 
exponential DTD with the same normalization~\citepalias[see figure 11 
of ][]{Weinberg2017}. 
\par 
Recently~\citet{Maoz2017} argued for a lower DTD normalization of 
$N_\text{Ia}/M_* = (1.3\pm0.1)\times10^{-3} M_\odot^{-1}$ for a Kroupa IMF, 
based on comparisons of the cosmic star formation history and the 
redshift-dependent SN Ia rate derived from cosmological surveys. Adopting this 
lower normalization would require us to adopt lower values of 
$y_\text{O}^\text{CC}$, $y_\text{Fe}^\text{CC}$, and $\eta$ to reproduce the 
observed [O/Fe]-[Fe/H] tracks in the Milky Way, reducing each by roughly a 
factor of two. Such a change is physically plausible, because many of the 
high-mass stars with the highest oxygen yields may collapse to black holes 
instead of exploding as CCSNe (see discussion by, e.g.,
~\citealp{Pejcha2015},~\citealp{Sukhbold2016}, and observational evidence 
of~\citealp{Gerke2015},~\citealp{Adams2017}). These changes would not alter our 
qualitative conclusions below, but they would change the detailed form of 
evolutionary tracks and element ratio distributions. \citet{Brown2019} found 
that the local specific SN Ia rate scales strongly (and inversely) with 
galaxy stellar mass, and they argue that this dependence may imply a 
metallicity-dependent $R_\text{Ia}(t)$ in addition to a DTD that produces more 
SNe Ia at early times. Adopting a metallicity-dependent $y_\text{Fe}^\text{Ia}$ 
would have a larger qualitative impact on our models (though as a practical 
matter it would be straight-forward to implement within \texttt{VICE}). We 
reserve a more thorough investigation of empirical constraints on elemental 
yields to future work. 
\par 
The AGB yields of s-process elements depend strongly on both stellar mass and
birth metallicity. It is therefore not feasible to specify single yield values 
or simple time-dependent functional forms. Instead, \texttt{VICE} implements 
a grid of fractional yields on a table of stellar mass and metallicity. At 
each timestep, and for each element, it then determines the appropriate yield 
$y_\text{x}^\text{AGB}$ in equation~\refp{eq:mdot_agb} via bilinear 
interpolation between elements on the grid. The current version of 
\texttt{VICE} allows users to adopt either the~\citet{Cristallo2011} 
or~\citet{Karakas2010} yield tables, and we adopt the former for calculations 
in this paper. A future version of \texttt{VICE} will likely include more yield 
tables as well as the capability to handle user-specifications on the AGB 
yields of each element. We provide further discussion of Sr yields in 
\S~\ref{sec:sr}. 

\section{Fiducial Starburst Models}
\label{sec:fiducial}

\begin{figure*} 
\includegraphics[scale = 0.31]{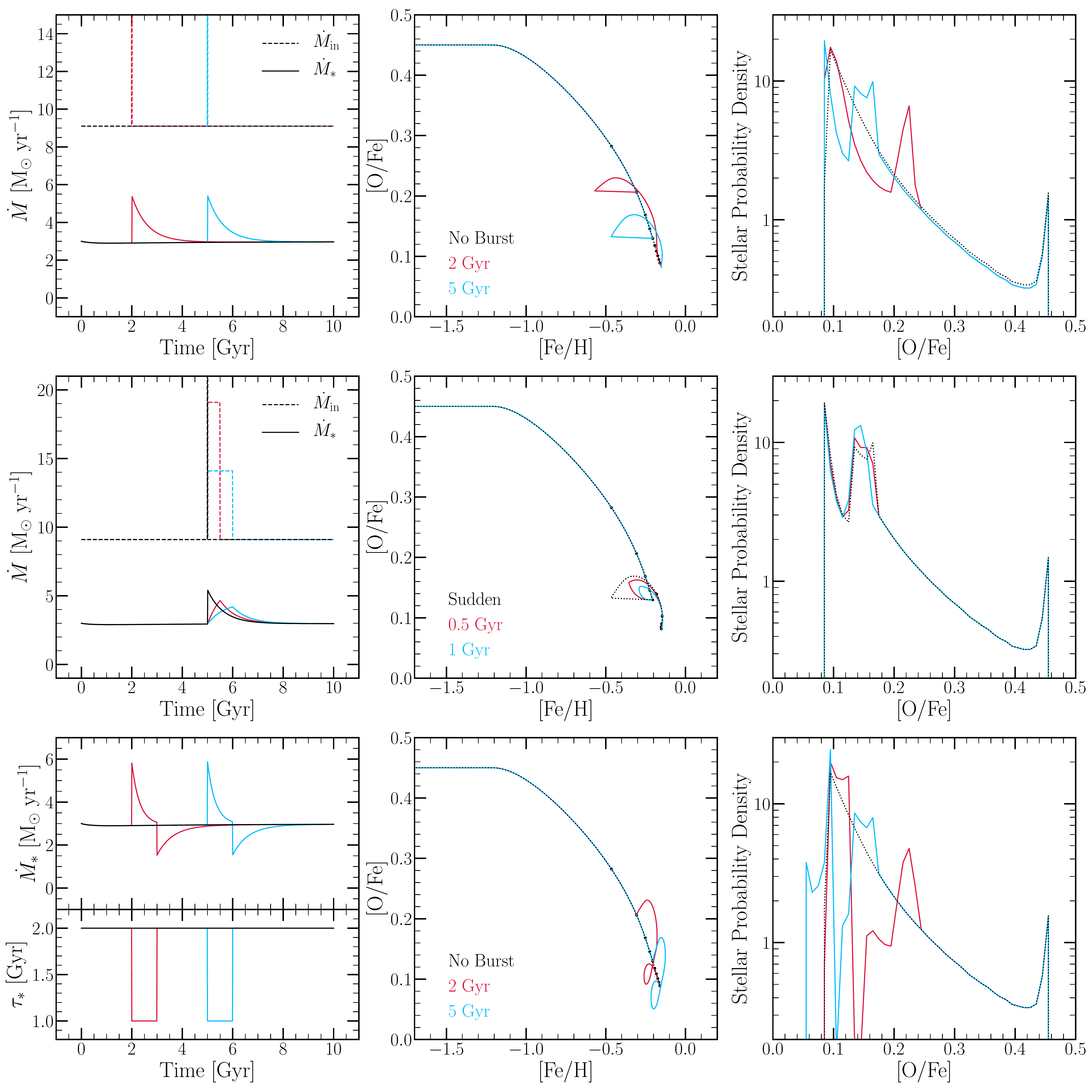}
\caption{
Evolutionary tracks in the [O/Fe]-[Fe/H] plane (middle column) and [O/Fe] 
distributions (right column) of starburst GCE models with infall and star 
formation histories shown in left panels. \textbf{Top}: Black curves show an 
unperturbed model with a constant infall rate and near-constant star formation 
rate. Red and blue curves show models with starbursts induced by adding 85\% of 
the ISM mass worth of $Z = 0$ gas at $t = 2$ or $t = 5$ Gyr, 
respectively. \textbf{Middle}: Red and blue curves show models in which the 
gas infall rate is boosted over a time interval of $\Delta t$ = 0.5 Gyr or 1 
Gyr, respectively, at $t = 5$ Gyr. Black curves show the sudden gas infall 
model from the upper row for comparison. The total amount of gas added is the 
same in all three models. \textbf{Bottom}: Red and blue curves show models with 
starbursts induced by doubling the SFE (halving $\tau_*$) for an interval of 
$\Delta t = 1$ Gyr at $t = 2$ Gyr or 5 Gyr, respectively, with the infall rate 
(not plotted) held constant. Black curves show the unperturbed model. In the 
middle panels, small points on the unperturbed model curve mark 1 Gyr 
intervals. 
}
\label{fig:fiducial_cases}
\end{figure*} 

\begin{figure*} 
\includegraphics[scale = 0.31]{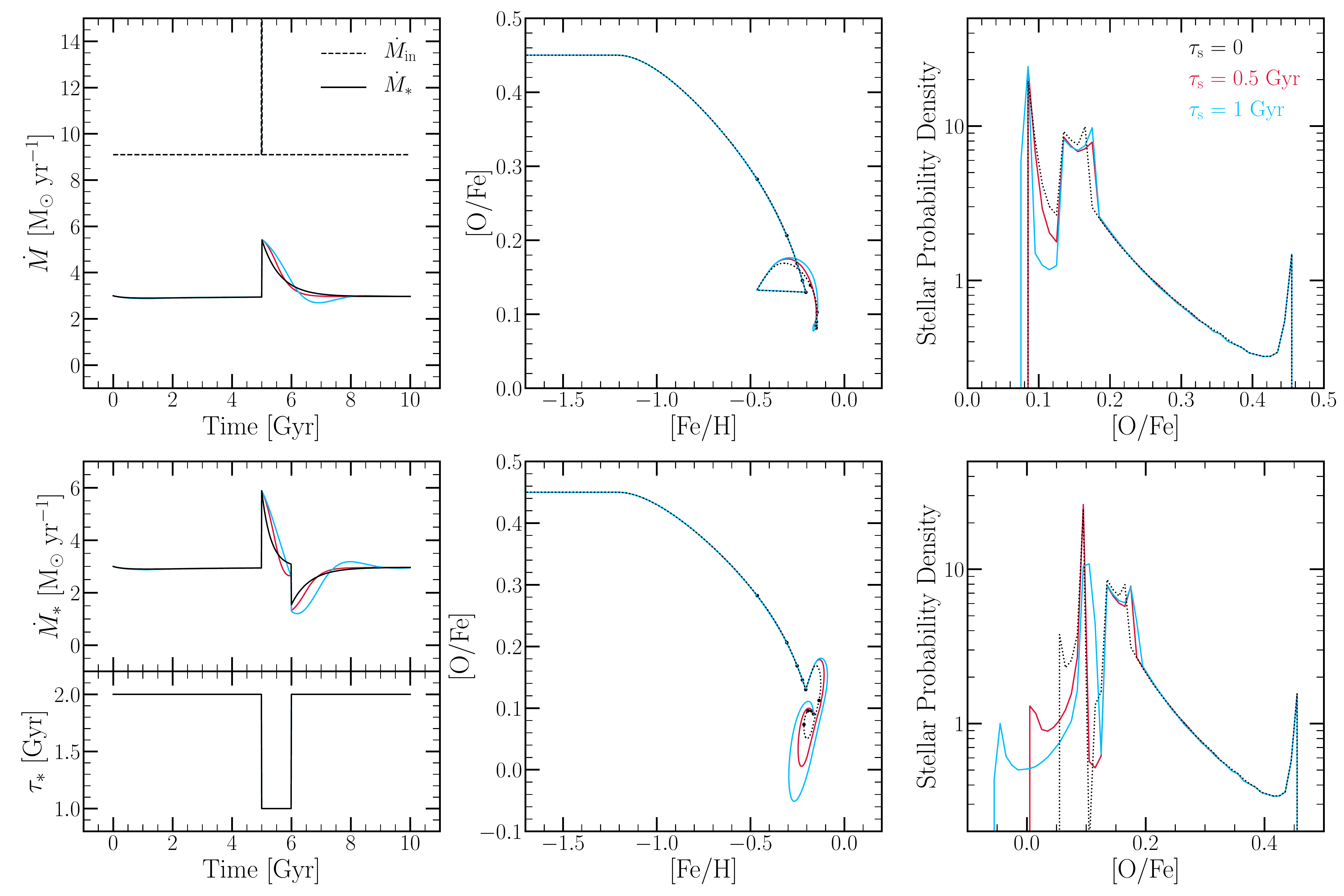}
\caption{
Similar to Fig.~\ref{fig:fiducial_cases}, for models in which the outflow 
$\dot{M}_\text{out} = \eta\langle\dot{M}_*\rangle_{\tau_\text{s}}$ responds 
to the SFR averaged over a preceding interval $\tau_\text{s} = 0.5$ Gyr (red) 
or 1 Gyr (blue). Top and bottom rows show models in which the starburst is 
induced by increasing the gas supply or SFE, respectively, at $t = 5$ Gyr, as 
in the top and bottom rows of Fig.~\ref{fig:fiducial_cases}. Black dotted 
curves show the corresponding $\tau_\text{s} = 0$ models, repeated from 
Fig.~\ref{fig:fiducial_cases}, with small dots at 1 Gyr intervals in the 
middle panels.  
}
\label{fig:ts_combined}
\end{figure*} 

We begin by defining a GCE model with nearly constant star formation, which we 
will then perturb in a variety of ways. Our fiducial no-burst model has an 
infall rate of $\dot{M}_\text{in} = 9.1\ M_\odot\ \text{yr}^{-1}$ onto a galaxy 
with an initial gas supply of $M_\text{g} = 6.0\times10^9\ M_\odot$, an SFE 
timescale of $\tau_*$ = 2 Gyr, a mass-loading factor of $\eta$ = 2.5, 
$\tau_\text{s}$ = 0, and $\xi_\text{enh}$ = 1 (i.e. Z$_\text{out}$ = 
Z$_\text{gas}$) with continuous recycling. We also adopt a power-law SN Ia 
delay-time distribution (DTD) with R$_\text{Ia} \propto t^{-1.1}$ and minimum 
delay time of $t_\text{D}$ = 150 Myr. In short, this is a model with a 
constant infall rate and (nearly) constant star formation rate with parameters 
that do not change with time. The analytic model of~\citetalias{Weinberg2017} 
accurately describes the [O/Fe] evolution of this numerical model. Although we 
adopt explicit numerical values for the initial gas mass and 
$\dot{M}_\text{in}$, the [Fe/H] and [O/Fe] evolution would be unchanged if we 
multiplied both of these quantities by the same constant factor. As shown 
by~\citetalias{Weinberg2017}, the characteristic time for the evolution of O 
or other CCSN elements in such a model is the depletion time 
$\tau_\text{dep} \equiv \tau_*/(1 + \eta - r_\text{inst})$, while for Fe the 
evolutionary timescale depends on both $\tau_\text{dep}$ and the characteristic 
SN Ia timescale $\tau_\text{Ia}\sim 1-2$ Gyr.  

\subsection{Gas-Driven Starbursts}
\label{sec:gas-driven}

\begin{figure*} 
\includegraphics[scale = 0.31]{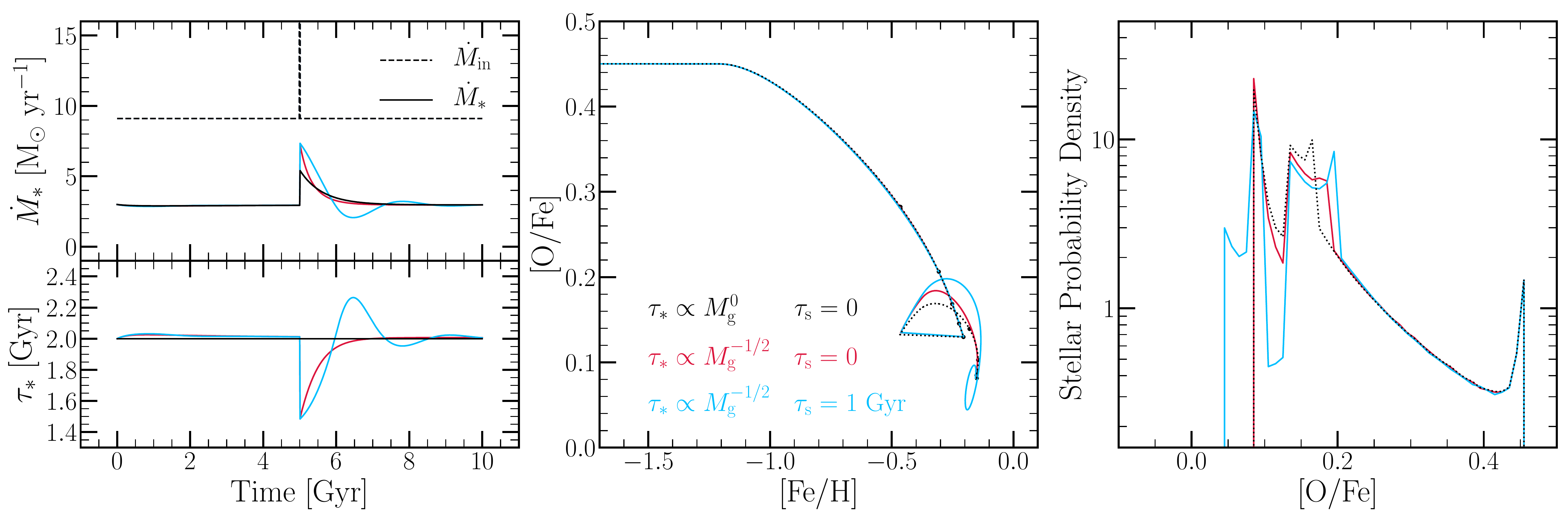}
\caption{
Similar to Fig.~\ref{fig:fiducial_cases}, for models in which the gas supply 
increases suddenly at $t = 5$ Gyr and the SFE timescale remains constant 
(black dotted) or decreases in accordance with the Kennicutt-Schmidt law (red, 
blue). The blue curve model has smoothing timescale $\tau_\text{s} = 1$ Gyr, 
and the black curve model is identical to the $t = 5$ Gyr starburst in the top 
row of Fig.~\ref{fig:fiducial_cases}. The lower left panel shows the response 
of $\tau_*$ to the evolving gas supply. 
}
\label{fig:ts_bolus_schmidt}
\end{figure*}

Our simplest starburst model is one in which a large amount of gas with a 
specified metallicity is added to the galaxy in a short amount of time. Here 
``large'' means that the added gas is significant compared to the current gas 
supply and ``short'' is relative to the timescales associated with GCE, in 
particular the depletion time $\tau_\text{dep}$. In this paper, we adopt the 
simplest scenario in which the added gas has zero metallicity, but any value 
can be used in \texttt{VICE}. 
\par
The top row of Fig.~\ref{fig:fiducial_cases} compares two gas-driven 
starburst models to our burstless scenario. These models have the same 
parameters as the burstless scenario with the exception of the infall rate. In 
these models, the infall rate assumes a value of $\dot{M}_\text{in}$ = 5000\ 
$M_\odot \text{yr}^{-1}$ for one $\Delta t$ = 1 Myr timestep, thereby adding 
$\dot{M}_\text{in}\Delta t = 5\times10^9 M_\odot$ of zero metallicity gas 
essentially instantaneously. Red and blue curves show models with gas added at 
$t = 2$ and $5$ Gyr, respectively. In each case, the nearly doubled gas supply 
causes a near doubling of the star formation rate (SFR). This burst decays on 
a timescale of~$\sim$1 Gyr as the excess gas is consumed by star formation and 
outflows. 
\par
The evolution of these models in [O/Fe] vs. [Fe/H] exhibits a ``jump-and-hook'' 
trajectory. Dilution by pristine gas causes an instantaneous jump to lower 
[Fe/H] at fixed [O/Fe]. The burst of star formation elevates the rate of CCSN 
enrichment to SN Ia enrichment, so the ISM evolves to higher [O/Fe] as the 
metallicity increases. Eventually the impact of the starburst dies away and 
the [O/Fe] evolution returns to that of the unperturbed model. 
\par 
The top right panel shows the normalized distribution of [O/Fe] in these 
models. The unperturbed model has two peaks in this distribution, the first at 
[O/Fe] $\approx$ +0.45 for stars formed early in the model galaxy's evolution 
when SN Ia enrichment is still negligible, and the second at 
[O/Fe] $\approx$ +0.08 produced when the system has reached equilibrium and is 
forming stars at constant [Fe/H] and [O/Fe]. For our adopted yields, a constant 
SFR model evolves to slightly super-solar [O/Fe], but a mildly declining SFR 
model would evolve to solar [O/Fe]~\citepalias[see][figure 3]{Weinberg2017}. A 
declining SFR would also boost the equilibrium [Fe/H] to solar instead of 
mildly sub-solar for our adopted yields and $\eta$. We have chosen to focus on 
perturbations of a constant SFR model for simplicity, but we have checked that 
our qualitative conclusions hold if the underlying model has exponentially 
declining star formation with $\tau_\text{sfh}\approx$ 6 Gyr. 
\par 
The starburst models produce a third peak in this distribution at intermediate 
values of [O/Fe]. The lower edge of this peak corresponds to the value of 
[O/Fe] at the start of the burst, and the upper edge corresponds to the value 
of [O/Fe] at the top of the hook seen in the middle panel. The peak arises both 
because the system spends extra time at these [O/Fe] values and because the 
SFR is elevated during this time. Although the jump-and-hook trajectories are 
similar for the two starburst models, the arc in [O/Fe] is flatter for the 
earlier burst, which corresponds to a narrower peak in [O/Fe]. This difference 
arises because at $t$ = 2 Gyr the CCSN/SN Ia ratio of the unperturbed model is 
still elevated compared to its eventual equilibrium ratio, so the extra boost 
from the starburst has a smaller relative impact. 
\par 
A gas rich merger or violent dynamical disturbance may induce a very rapid 
increase in a galaxy's supply of star-forming gas. However, a temporary boost 
in a galaxy's gas accretion rate can also induce elevated star formation. 
The middle row of Fig.~\ref{fig:fiducial_cases} compares the $t = 5$ Gyr 
instantaneous gas increase model to models in which the same 
$5\times10^9 M_\odot$ of gas is added over $\Delta t$ = 0.5 and 1.0 Gyr 
intervals. The perturbation to the SFR is smoother (left panel), though the 
number of ``extra'' stars formed is similar. The hooks in [O/Fe]-[Fe/H] are no 
longer flat-bottomed because the elevated SFR increases [O/Fe] at the same 
time that the infall dilutes [Fe/H]. For $\Delta t$ = 1 Gyr the jump in [O/Fe] 
is small because the maximum boost of SFR is only about half that of the 
instantaneous model. However, the extra peak in the [O/Fe] distribution is 
remarkably similar in all three models, though slightly sharper for 
$\Delta t$ = 1 Gyr. 
\par 
We conclude that a third peak in the [O/Fe] distribution is the characteristic 
observable signature of a gas-driven starburst that formed a significant 
fraction of a system's stars. The location of the peak indicates the value of 
[O/Fe] at the time of the burst. Resolving these peaks requires a large sample 
of stars with precise [O/Fe] (or [$\alpha$/Fe]) values, i.e. statistical errors 
of 0.05 dex or below. Correlating these [$\alpha$/Fe] measurements with 
individual stellar age estimates could increase the diagnostic power even if 
the age estimates have substantial statistical errors.

\subsection{Efficiency-Driven Starbursts}
\label{sec:efficiency-driven}
The bottom row of Fig.~\ref{fig:fiducial_cases} shows a scenario in which 
starbursts arise from a temporary increase of SFE instead of an increase in 
gas supply. We double the SFE -- thus decreasing the SFE timescale $\tau_*$ 
from 2 Gyr to 1 Gyr -- for a period of $\Delta t$ = 1 Gyr, beginning at $t = 2$ 
or 5 Gyr. The gas infall rate is held constant. As in the gas-doubling 
scenario, the SFR initially jumps by a factor of two, then decays to its 
original value. However, once $\tau_*$ returns to 2 Gyr, the SFR drops 
\textit{below} that of the unperturbed model because the gas supply has been 
depleted during the high SFE phase. Over an interval of~$\sim$1 Gyr, the SFR 
recovers to the value of $\dot{M}_*\approx3\ M_\odot\ \text{yr}^{-1}$ at which 
star formation and outflow balance infall. 
\par 
The hooks in [O/Fe]-[Fe/H] evolution have a different morphology for the 
efficiency-driven bursts. Because there is no dilution by low metallicity gas, 
the tracks jump up to higher [O/Fe] with slightly increasing [Fe/H], instead 
of first moving back to lower [Fe/H]. Furthermore, because of the depression 
in SFR once $\tau_*$ returns to its baseline value, the [O/Fe] track dips 
below that of the unperturbed model before returning to it at late 
times. During the downward loop, the rate of SNe Ia is high because of the 
stars formed during the recent burst, but the rate of CCSNe is low due to the 
reduced SFR. 
\par 
The distribution of [O/Fe] in these models again shows an extra peak at [O/Fe] 
values close to those at the onset of the starburst. However, the morphology 
of these distributions is different from that of the gas-driven starburst 
models in two ways. First, the peak in [O/Fe] is followed by a much deeper 
trough at slightly lower [O/Fe] because the SFR is depressed while the ISM is 
evolving through this abundance ratio. Second, the [O/Fe] distribution 
acquires an additional peak at a value that corresponds to the bottom of the 
downward loops in the middle panel. With sufficiently good data, it might be 
possible to distinguish the signature of gas-driven and efficiency-driven 
starbursts from the detailed shape of the [O/Fe] distribution. In particular, 
an efficiency-driven burst at relatively late times would produce a population 
of roughly coeval stars with [$\alpha$/Fe] values below that of the bulk 
population. There is some hint of such a population in the solar 
neighborhood~\citep{Feuillet2018}. 

\subsection{Outflow Smoothing Time}
\label{sec:smoothing}

We now examine models in which the outflow rate $\dot{M}_\text{out}$ responds 
to the SFR averaged over a time interval $\tau_\text{s}$ instead of the 
instantaneous SFR. Fig.~\ref{fig:ts_combined} shows star formation histories, 
[O/Fe]-[Fe/H] tracks, and [O/Fe] distributions for gas-driven and 
efficiency-driven starburst models with $\tau_\text{s}$ = 0, 0.5, and 1 Gyr. 
The $\tau_\text{s}$ = 0 models are identical to the t = 5 Gyr burst models 
shown in the top and bottom rows of Fig.~\ref{fig:fiducial_cases}. Because 
the enhanced infall models (middle row of Fig.~\ref{fig:fiducial_cases}) are 
qualitatively similar to the instantaneous gas doubling model (top row), we 
show only this limiting case of a gas-driven starburst in the remainder of the 
paper. 
\par 
For the gas-driven starburst, even a 1 Gyr smoothing time has only a small 
impact on the [O/Fe]-[Fe/H] trajectory and [O/Fe] distribution. Just after the 
accretion event, the SFR in the smoothed models is slightly higher than in the 
$\tau_\text{s}$ = 0 model because the outflow rate is lower, and the hook in 
the evolutionary track therefore reaches slightly higher [O/Fe]. For 
$\tau_\text{s}$ = 1 Gyr, the SFR at $t\approx$ 6 - 8 Gyr dips below the 
3 $M_\odot\ \text{yr}^{-1}$ baseline, because the extra accreted gas has been 
consumed and the outflow rate remains high because of the earlier starburst. 
As a result, the deficit in the [O/Fe] distribution at [O/Fe] $\approx$ +0.1 
is deeper in this model. 
\par 
For the efficiency-driven starburst, smoothing has a larger impact because the 
delayed outflow deepens the depression of SFR after the burst. The downward 
hook of [O/Fe] is therefore substantially deeper even for $\tau_\text{s}$ = 0.5 
Gyr. Smoothing of the outflow response exaggerates the characteristic form of 
an efficiency-driven starburst perturbation and moves the extra peak of the 
[O/Fe] distribution to a lower value.

\subsection{Hybrid Starbursts}
\label{sec:hybrid}

If the gas supply of a galaxy increases suddenly, then the SFE may also 
increase because of greater gas self-gravity, more rapid cloud collisions, or 
whatever dynamical disturbance drove the gas increase in the first place. 
Observations provide some evidence for starbursts that are driven by both 
increased gas supply and increased 
SFE~\citep[][and the citations therein]{Kennicutt2012}. 
Fig.~\ref{fig:ts_bolus_schmidt} shows results for a hybrid model in which a 
doubling of the gas supply is linked to a Kennicutt-Schmidt scaling of the SFE, 
with $\tau_* = (2\text{ Gyr})(M_\text{g}/6\times10^9\ M_\odot)^{-1/2}$ (see 
\S~\ref{sec:methods}). If the smoothing time $\tau_\text{s}$ = 0, then the 
evolution of this hybrid model is only slightly different from that of our 
standard gas-driven starburst, as one can see by comparing the dashed black and 
solid red curves in the middle and right panels of 
Fig.~\ref{fig:ts_bolus_schmidt}. The hybrid burst has a higher peak SFR, 
which leads to a higher peak of the [O/Fe] hook. For $\tau_\text{s}$ = 1 Gyr, 
the trajectory and [O/Fe] distribution of the hybrid model show features of 
both the gas-driven and efficiency-driven models. In particular, this model 
has a period of depressed SFR because of the delayed ejection of gas by the 
starburst, and the enhanced ratio of SNe Ia/CCSNe during this period causes a 
downward hook in [O/Fe] and an additional peak in the [O/Fe] distribution. 

\section{Strontium}
\label{sec:sr}
\subsection{Nucleosynthesis}
\label{sec:sr_nuc}

\begin{figure} 
\includegraphics[scale = 0.45]{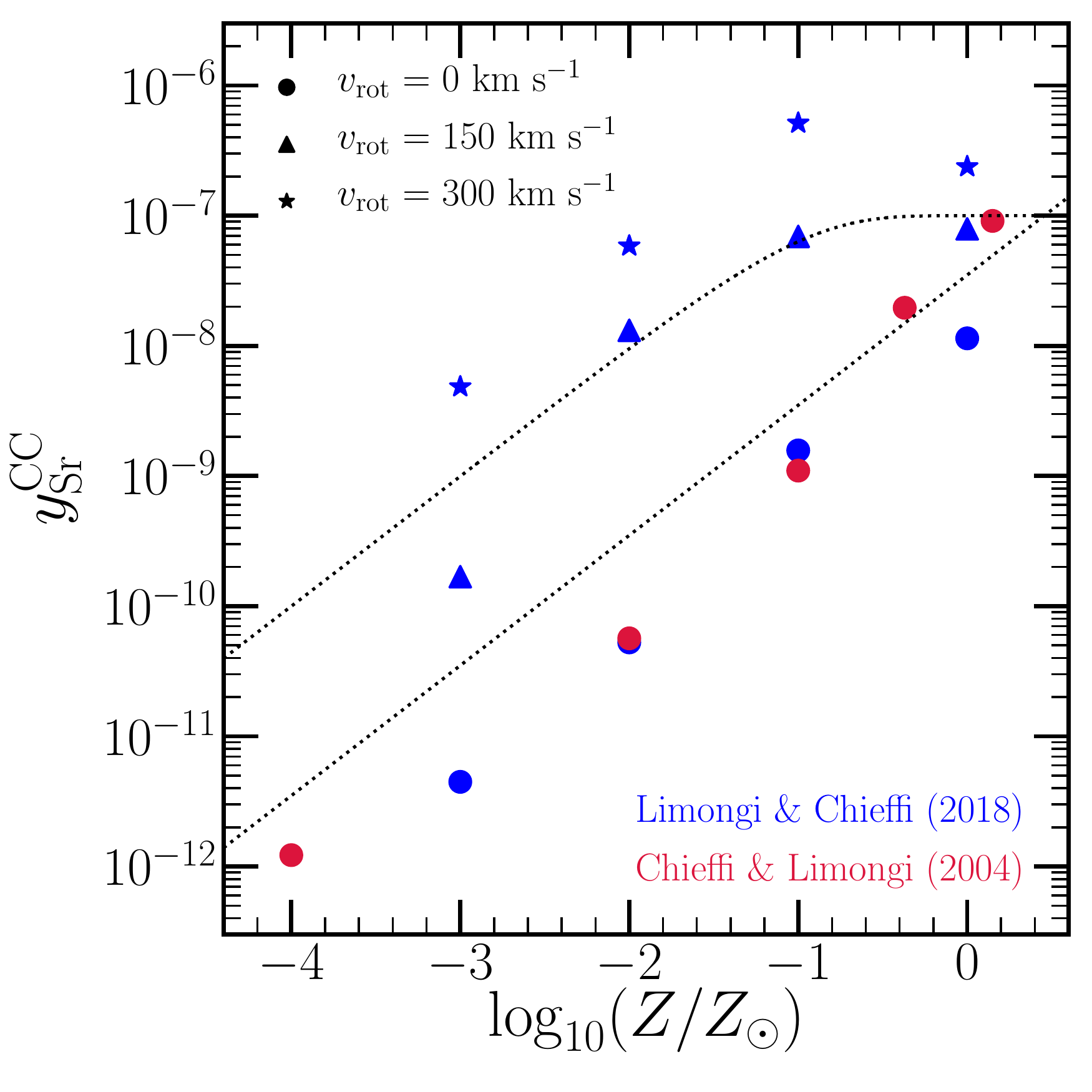}
\caption{
IMF-averaged CCSN yields of Sr computed using the non-rotating progenitor 
models of (\citealp{Chieffi2004}, red circles) and the models 
of~\citet{Limongi2018} for progenitors with $v_\text{rot} = 0$ (blue circles), 
150 km s$^{-1}$ (blue triangles), and 300 km s$^{-1}$ (blue stars). Dotted 
curves show approximate characterizations of these results used in our GCE 
models, $y_\text{Sr}^\text{CC} = 3.5\times10^{-8}(Z/Z_\odot)$ and 
$y_\text{Sr}^\text{CC} = 10^{-7}(1 - \exp{(-10Z/Z_\odot)})$. We adopt $Z_\odot$ = 
0.014 based on~\citet{Asplund2009}. 
}
\label{fig:sr_cc_yields}
\end{figure} 

\begin{figure*} 
\includegraphics[scale = 0.32]{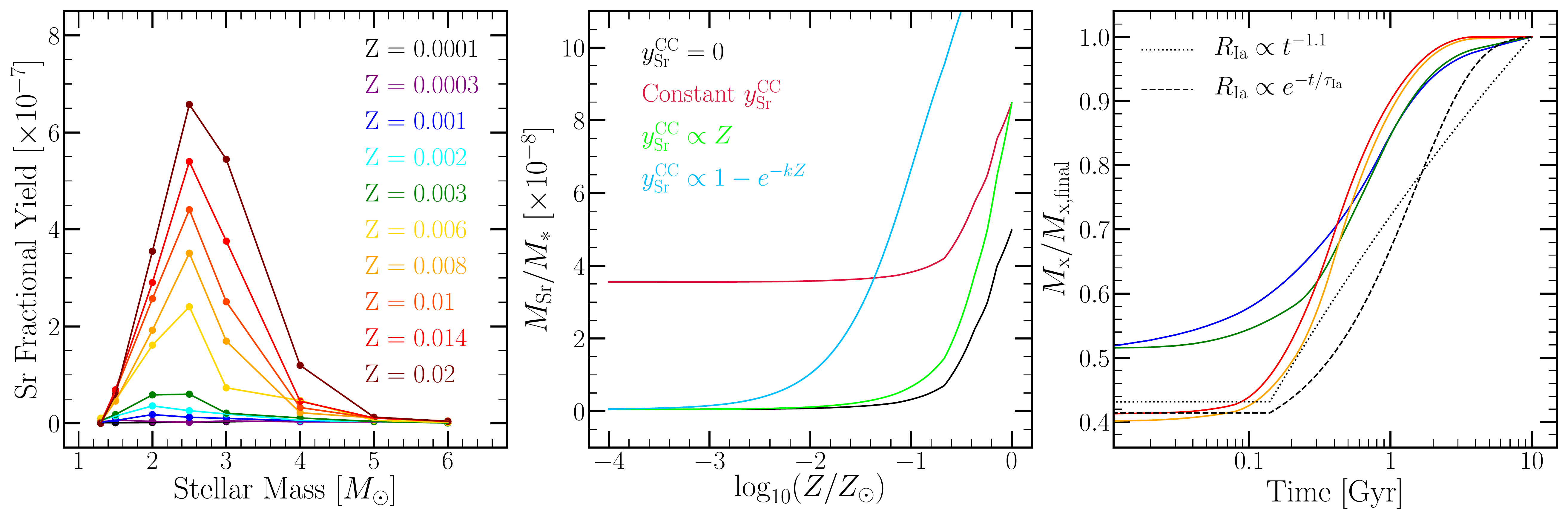}
\caption{
Yields of Sr as a function of stellar mass, metallicity, and time. 
\textbf{Left}: Fractional yields of AGB stars - the ejected Sr mass divided by 
the initial stellar mass, computed as a function of stellar mass and 
metallicity using the FRANEC code~\citep{Cristallo2011}. \textbf{Middle}: 
IMF-averaged Sr yield after 10 Gyr, for a single stellar population of 
metallicity $Z$ formed at $t = 0$, computed by adding the AGB yields to CCSN 
yields illustrated by the dotted curves in Fig.~\ref{fig:sr_cc_yields}, or to 
constant yields $y_\text{Sr}^\text{CC} = 3.5\times10^{-8}$ or 
$y_\text{Sr}^\text{CC} = 0$ (AGB only). \textbf{Right}: Time evolution of Sr 
production for single stellar populations of metallicity $Z$ = 0.001, 0.003, 
0.008, and 0.014, assuming the $y_\text{Sr}^\text{CC} \propto Z$ model. 
Curves are color-coded to the legend in the left panel. All curves are 
normalized to the final Sr mass produced after 10 Gyr, which depends strongly 
on $Z$ as shown in the right panel. Dotted and dashed black curves show the 
time evolution of Fe for our standard values of $y_\text{Fe}^\text{CC}$ and 
$y_\text{Fe}^\text{Ia}$ and a $t^{-1.1}$ or $e^{-t/1.5\text{ Gyr}}$ DTD with a 
minimum delay time of 0.15 Gyr. Because AGB production is dominated by 
$2-4\ M_\odot$ stars, AGB Sr enrichment from a single stellar population 
occurs faster than SN Ia Fe enrichment.  
}
\label{fig:sryields_3panel}
\end{figure*}

Strontium is one of the commonly used tracers of s-process nucleosynthesis in 
AGB stars~\citep[e.g.][]{Conroy2013,Mishenina2019}. Sr production differs from 
that of O and Fe, the two elements that we have examined thus far, because the 
delay time of AGB enrichment differs from that of SNe Ia and because the Sr 
yields of both CCSNe and AGB stars are expected to depend strongly on 
metallicity. Both of these differences have an important impact on predicted 
evolutionary tracks and element ratio distributions. 
\par 
Fig.~\ref{fig:sr_cc_yields} plots IMF-averaged net CCSN yields of strontium 
based on the models of~\citet{Chieffi2004} and~\citet{Limongi2018}. These 
yields are defined by: 
\begin{equation} 
\label{eq:frac_yield}
y_x^\text{CC} = \ddfrac{
	\int_8^u m_x \frac{dN}{dm}dm 
}{ 
	\int_l^u m \frac{dN}{dm}dm 
}
\end{equation} 
where $m_x$ is the mass of the element $x$ ejected in the explosion of a star 
of mass $m$, and $dN/dm$ is the assumed stellar IMF, for which we 
adopt~\citet{Kroupa2001}. We adopt $l$ = 0.08 $M_\odot$ and $u$ = 100 $M_\odot$ 
as the lower and upper mass limits of the IMF and 8 $M_\odot$ as the minimum 
progenitor mass for a CCSN explosion. 
\par 
In practice, supernova nucleosynthesis studies determine the value of 
$m_\text{x}$ for of order 10 values of $m$ at a specified metallicity and 
rotational velocity. To compute the numerator of equation~\refp{eq:frac_yield}, 
\texttt{VICE} linearly interpolates $m_\text{x}$\footnote{
	linearly in $m$, not $\log m$
} values between the two surrounding $m$ values in the available yield 
grid, or linearly extrapolates $m_\text{x}$ values from the two highest $m$ 
values in the grid if it does not extend to 100 $M_\odot$. 
\par 
\citet{Chieffi2004} report Sr yields for non-rotating CCSN progenitors 
($v_\text{rot}$ = 0) at a wide range of metallicities, while \citet{Chieffi2013} 
report yields for $v_\text{rot}$ = 0 and 300 km s$^{-1}$ but at only solar 
metallicity. Progenitor rotation affects Sr yields from CCSNe due to 
rotationally induced mixing~\citep{Frischknecht2016}. We presume the results 
of~\citet{Chieffi2013} to be superseded by those of~\citet{Limongi2018}, who 
examined a range of metallicites and values of $v_\text{rot}$ = 0, 150 km 
s$^{-1}$, and 300 km s$^{-1}$. However, we caution that the impact of rotation 
on the Sr yield at solar metallicity is much stronger in 
the~\citet{Limongi2018} study than in~\citet{Chieffi2013} due to a different 
calibration of the rotation-induced mixing efficiency. 
\par 
Fig.~\ref{fig:sr_cc_yields} shows that the predicted CCSN yields depend 
strongly on metallicity and are much higher (typically 1-3 orders of magnitude) 
for rapidly rotating vs. non-rotating progenitors. As approximate descriptions 
of the numerical results, we show the functions 
\begin{subequations}\begin{align} 
y_\text{Sr}^\text{CC} &= 3.5\times10^{-8}(Z / Z_\odot) 
\label{eq:y_sr_cc_linear} \\ 
\intertext{for v$_\text{rot}$ = 0 and} 
y_\text{Sr}^\text{CC} &= 10^{-7}\left[1 - e^{-10(Z/Z_\odot)}\right] 
\label{eq:y_sr_cc_limexp} 
\end{align}\end{subequations} 
for $v_\text{rot}$ = 150 km s$^{-1}$. For comparison, we will also compute GCE 
models with a constant $y_\text{Sr}^\text{CC} = 3.5\times10^{-8}$ matched to 
our linear model at $Z = Z_\odot$ and with $y_\text{Sr}^\text{CC} = 0$ 
corresponding to pure AGB enrichment. We caution that these are not fits to 
the yields plotted in Fig.~\ref{fig:sr_cc_yields}; we adopt them as an 
agnostic approach to the form of the metallicity-dependent yield in the 
interval -2 $\lesssim$ [Fe/H] $\lesssim$ 0 in which our models are focused. 
\par
For AGB production of Sr, we use fractional yields as a function of initial 
stellar mass at various metallicities from the FRANEC 
code~\citep{Cristallo2011}. These are plotted in the left-hand panel of 
Fig.~\ref{fig:sryields_3panel}, and they show two notable features. First, 
for near-solar metallicity the fractional yields are sharply peaked at stellar 
masses of 2-3 $M_\odot$. To obtain the total mass yield per star one 
multiplies by $M$, giving weight to the contribution of higher mass 
stars, but the number of stars per linear $\Delta M$ interval is proportional 
to $M^{-2.3}$ for a Kroupa IMF in this mass range, thus increasing the weight 
of lower mass stars. The strong mass dependence of the fractional yields means 
that the IMF-averaged AGB yield is dominated by stars with relatively short 
lifetimes. The second notable feature is a strong metallicity dependence, 
expected because the amount of Sr produced via the s-process during the AGB 
phase should increase with the abundance of free neutrons produced by nuclear 
reactions involving C and Ne isotopes. For $Z\lesssim Z_\odot/3$ the predicted 
fractional yields are below $10^{-7}$ at all masses, and for 
$Z\gtrsim Z_\odot/3$ the maximum fractional yield is roughly proportional to 
$Z$. 
\par 
In the middle panel of Fig.~\ref{fig:sryields_3panel}, the black curve shows 
the late-time ($t = 10$ Gyr), IMF-averaged AGB Sr yield as a function of 
metallicity. At $Z = Z_\odot$, the yield is $y_\text{Sr}^\text{AGB} = 
5\times10^{-8}$, but for $Z < Z_\odot/3$ the yield is well below $10^{-8}$. 
The green curve shows the total yield from adding $y_\text{Sr}^\text{AGB}$ to 
the $y_\text{Sr}^\text{CC}$ of equation~\refp{eq:y_sr_cc_linear}, which 
approximates the non-rotating~\citet{Limongi2018} models. At all metallicities 
for which $y_\text{Sr} > 10^{-8}$, the CCSN and AGB contributions are 
comparably important. However, for the $v_\text{rot} = 150\text{ km s}^{-1}$ 
yields approximated by equation~\refp{eq:y_sr_cc_limexp}, the CCSN yields 
dominate over the AGB yields at all metallicities (blue curve). The red curve 
shows the simple case of adding $y_\text{Sr}^\text{AGB}$ to a constant 
$y_\text{Sr}^\text{CC} = 3.5\times10^{-8}$. 

\begin{figure*} 
\includegraphics[scale = 0.45]{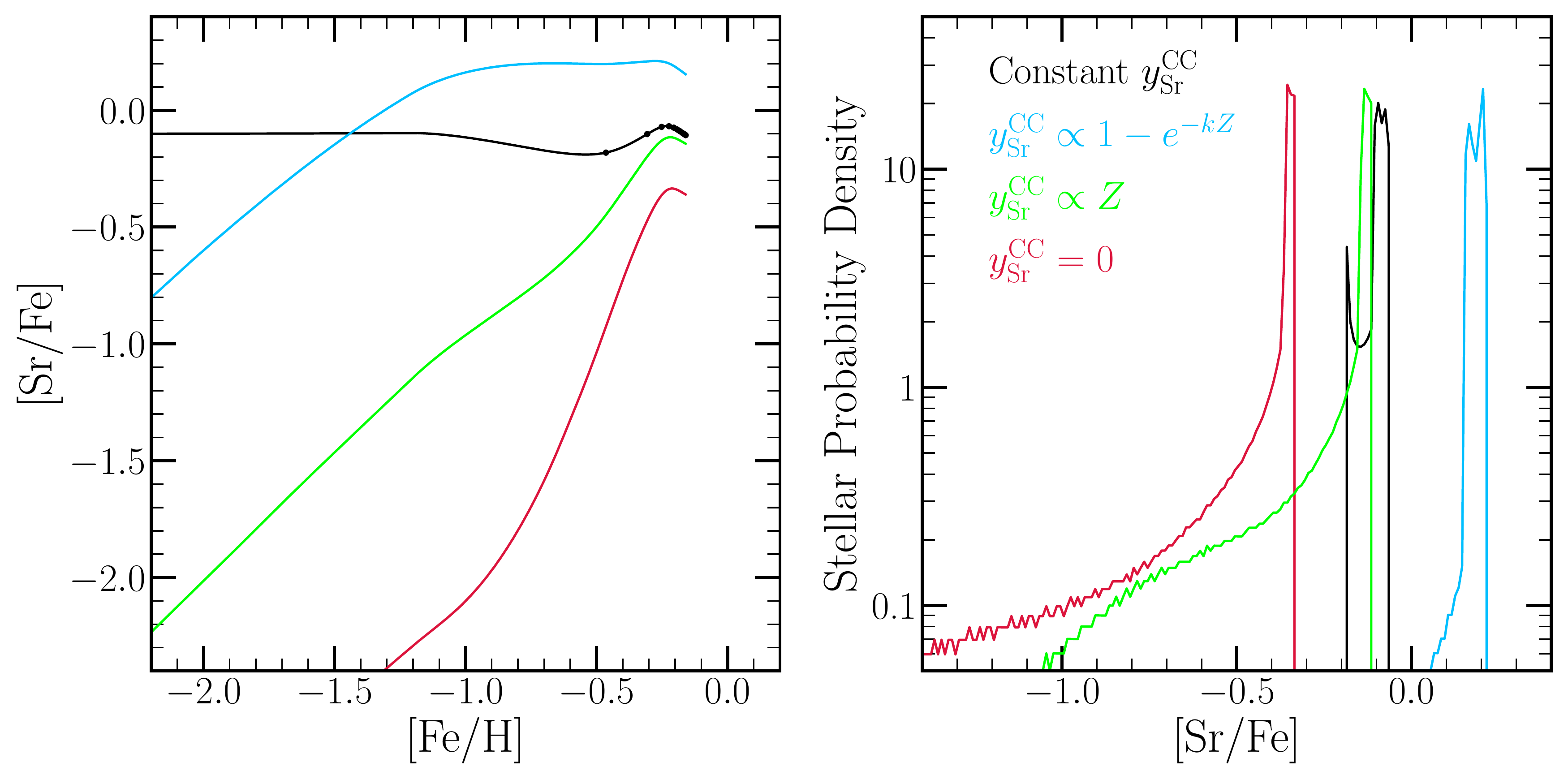}
\caption{
Evolutionary tracks (left) and final [Sr/Fe] distributions (right) under our 
fiducial burstless GCE model for four different assumptions about 
$y_\text{Sr}^\text{CC}$: constant yield, zero yield, and the 
metallicity-dependent yields for non-rotation or rotation progenitors 
described by equations~\refp{eq:y_sr_cc_linear} and~\refp{eq:y_sr_cc_limexp}. 
The red ($y_\text{Sr}^\text{CC} = 0$) curve shows the predicted evolution for 
our metallicity dependent AGB yields (\citealp{Cristallo2011}; 
Fig.~\ref{fig:sryields_3panel}), which are adopted in all four models. On the 
black curve, small black points are plotted at $\Delta t$ = 1 Gyr intervals, 
and all models reach a given [Fe/H] at the same time. 
}
\label{fig:sr_yields}
\end{figure*}

The right panel of Fig.~\ref{fig:sryields_3panel} shows the time evolution of 
Sr production for a selection of metallicity values shown in the left panel, 
$Z$ = 0.001, 0.003, 0.008, and 0.014. All curves are normalized by the 
late-time yield, which is strongly dependent on metallicity as shown in the 
middle panel. Here we adopt the $y_\text{Sr}^\text{CC} \propto Z$ yield model 
for CCSNe, and in all cases this accounts for about 40-50\% of the total yield. 
Typically about half of the AGB contribution comes within the first 0.5 Gyr, 
and nearly all of it within 2 Gyr. Dotted and dashed curves show the evolution 
of Fe production for our fiducial values of $y_\text{Fe}^\text{CC} = 0.0012$ 
and $y_\text{Fe}^\text{Ia} = 0.0017$ and a $t^{-1.1}$ DTD or an 
$e^{-t/1.5\text{ Gyr}}$ DTD, respectively. Although our assumed minimum delay 
is $t_\text{D} = 0.15$ Gyr, getting half of the SN Ia Fe contribution 
takes~$\sim0.9 - 1$ Gyr, so the AGB Sr enrichment is faster, albeit moderately, 
than the SN Ia Fe enrichment. This rapid AGB contribution is a consequence of 
the dominant contribution from $2 - 4\ M_\odot$ stars, which have short 
lifetimes. These curves represent the Sr production from a single population of 
stars at a given metallicity. In a GCE model the metallicity itself rises with 
time, thus increasing the Sr production because of the metallicity-dependent 
yield. In the next section, we demonstrate that this complicates the 
enrichment timescale of Sr relative to Fe. 

\begin{figure*} 
\includegraphics[scale = 0.31]{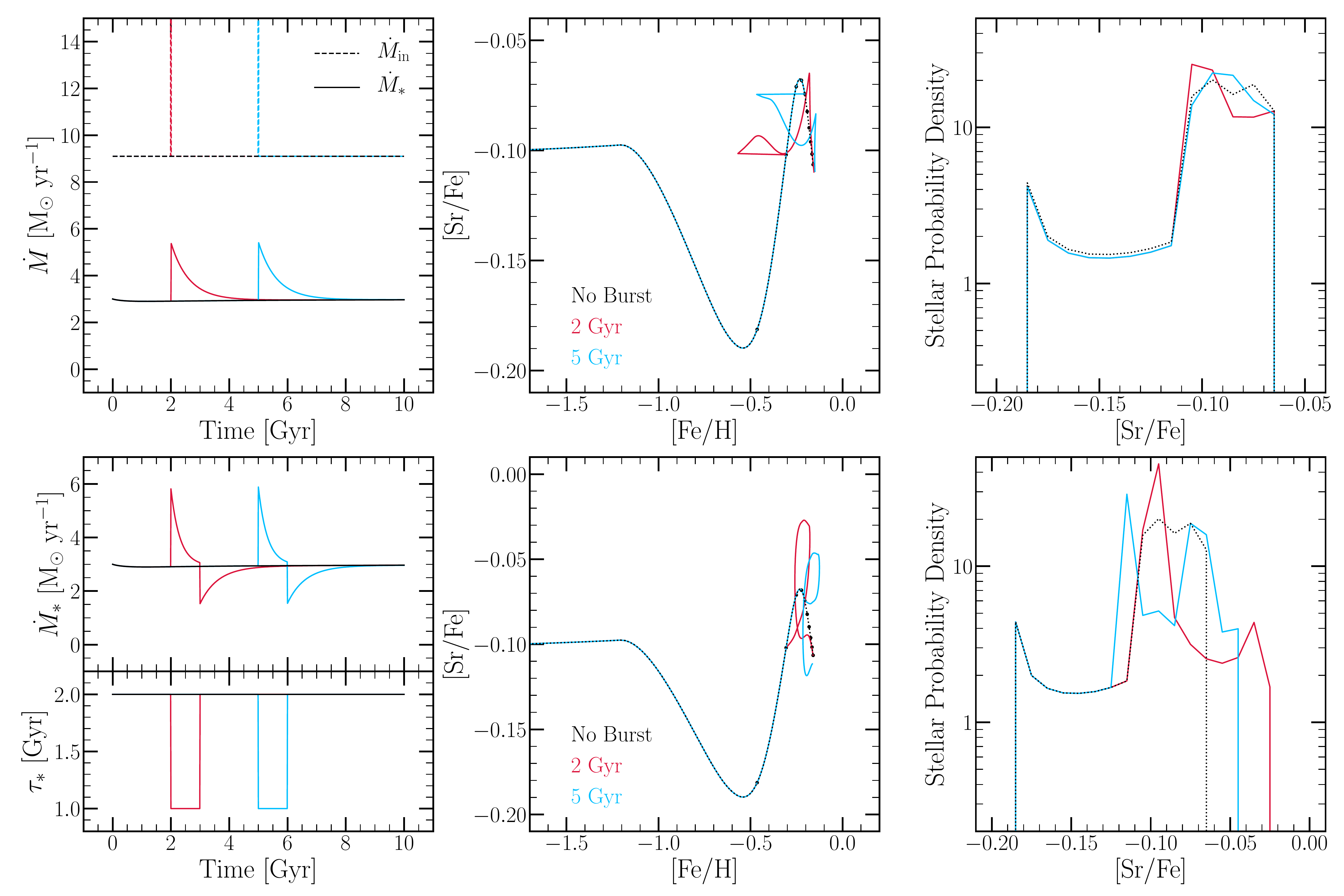}
\caption{
Evolutionary tracks (middle) and final [Sr/Fe] distributions (right) for our 
fiducial starburst models, analogous to the top and bottom rows of 
Fig.~\ref{fig:fiducial_cases}. All models adopt $y_\text{Sr}^\text{CC} = 
3.5\times10^{-8}$ and the~\citet{Cristallo2011} AGB yields illustrated in 
Fig.~\ref{fig:sryields_3panel}. 
}
\label{fig:bursts_srfe}
\end{figure*}

\subsection{Smooth Evolution}
\label{sec:sr_enrich} 

Fig.~\ref{fig:sr_yields} shows the [Sr/Fe]-[Fe/H] tracks and [Sr/Fe] 
distributions for our fiducial unperturbed GCE model, with constant SFR, 
$\tau_* = 2$ Gyr,~$\eta = 2.5$, and the AGB Sr yields illustrated in 
Fig.~\ref{fig:sryields_3panel}. We consider four different assumptions about 
CCSN yields: $y_\text{Sr}^\text{CC} = 0$, $y_\text{Sr}^\text{CC} = 
3.5\times10^{-8}$, and the metallicity dependent yields of 
equations~\refp{eq:y_sr_cc_linear} and~\refp{eq:y_sr_cc_limexp}. The 
$y_\text{Sr}^\text{CC} = 3.5\times10^{-8}$ model has a flat plateau at 
[Sr/Fe] = -0.1 for [Fe/H] $< -1.0$, which reflects the ratio of our constant 
CCSN yields. The [Sr/Fe] ratio then dips downward as SN Ia Fe enrichment 
becomes important, analogous to the knee in [O/Fe]-[Fe/H] evolution. However, 
as [Fe/H] rises further, AGB enrichment becomes competitive with CCSN 
enrichment, and [Sr/Fe] moves upward. After reaching a maximum at [Fe/H] = 
-0.2, [Sr/Fe] = -0.1, the [Sr/Fe] curves turns downward again because the 
timescale of SN Ia enrichment is longer than that of AGB enrichment. Even 
though single stellar populations generally produce Sr before Fe (see 
Fig.~\ref{fig:sryields_3panel}), the bulk of the Sr production in GCE follows 
the bulk production of more abundant elements like O and Fe due to the 
metallicity dependence of the yields. The [Sr/Fe] distribution of this model 
has a peak at [Sr/Fe]$\approx$-0.18, corresponding to the minimum in the 
[Sr/Fe]-[Fe/H] curve, and a second, higher peak at [Sr/Fe]$\approx$-0.1. In 
detail, this second peak is split in two, corresponding to the maximum in the 
[Sr/Fe]-[Fe/H] track and the slightly lower final equilibrium. 
\par 
With $y_\text{Sr}^\text{CC} = 0$ (AGB only), the [Sr/Fe] ratio is below -2 for 
[Fe/H] $<$ -1 and rises steeply with increasing [Fe/H], reaching a maximum at 
[Sr/Fe]$\approx$-0.35. Adding CCSN enrichment with $y_\text{Sr}^\text{CC} 
\propto Z$, corresponding approximately to the non-rotating~\citet{Limongi2018} 
yields, gives a shallower but still steeply rising [Sr/Fe]-[Fe/H] trend, which 
peaks at [Sr/Fe]$\approx$-0.1. Although one can see the imprint of SN Ia Fe 
enrichment on both of these curves, it is subtle relative to the strong trend 
arising from metallicity-dependent CCSN yields. 
\par 
Our approximate model of the rotating~\citet{Limongi2018} yields given by 
equation~\refp{eq:y_sr_cc_limexp} produces a [Sr/Fe] curve that rises rapidly 
until [Fe/H] = -1, then stays nearly constant at [Sr/Fe]$\approx$+0.2. AGB 
enrichment is small relative to CCSN enrichment in this model, as shown in 
Fig.~\ref{fig:sryields_3panel}. There is still a slight dip in [Sr/Fe] at 
late times, producing a split in the [Sr/Fe] distribution. 
\par 
Spectra of early-type galaxies imply [Sr/Fe]$\approx$0 for stellar populations 
typically dominated by solar or mildly super-solar 
metallicities~\citep{Conroy2013}, showing that solar abundance ratios arise 
even in systems with very different star formation histories from the Milky 
Way. Measurements of individual stars in the Milky Way and in dwarf satellites 
show median trends that are roughly flat at [Sr/Fe]$\approx$0 down to 
[Fe/H]$\approx$-3, though the star-to-star scatter becomes large below 
[Fe/H] = -1~\citep[see, e.g.,][and references therein]{Mishenina2019, 
Hirai2019}. Above [Fe/H] = -1, our model with the~\citet{Limongi2018} rotating 
CCSN progenitor yields produces a flat [Sr/Fe] trend, but only our 
$y_\text{Sr}^\text{CC}$ = constant model produces a flat trend to [Fe/H] as low 
as -3. We conclude that reproducing Milky Way observations requires an 
additional source of Sr that is prompt compared to SN Ia enrichment and 
approximately independent of metallicity at least for [Fe/H] $< -1$. This is 
in agreement with more detailed models of Sr enrichment investigating a 
variety of potential sources, such as neutron star mergers, electron-capture 
and magnetorotationally driven supernovae, and rotating massive 
stars~\citep[e.g.][]{Cescutti2014, Cescutti2015, Prantzos2018, Hirai2019, 
Rizzuti2019}. Neutron-rich neutrino-driven winds from newly 
formed neutron stars should also produce Sr via r-process nucleosynthesis in 
core collapse supernovae~\citep{Thompson2001,Vlasov2017,Thompson2018}, and 
this production is typically not included in calculations of CCSN yields such 
as~\citet{Limongi2018}. Sources that produce relatively large amounts of Sr in 
events that are individually rare would help to explain the large star-to-star 
scatter at low [Fe/H]. We conclude that our constant $y_\text{Sr}^\text{CC} = 
3.5\times10^{-8}$ model could retroactively account for this contribution; 
this arises from the nature of equation~\refp{eq:mdot_ccsne} which in principle 
could fold all prompt enrichment components into $y_\text{Sr}^\text{CC}$ as a 
function of metallicity. Nonetheless, we encourage caution that sufficiently 
accurate modeling of Sr production at metallicities as low as [Fe/H] $< -2$ 
may require a more complete understanding of the astrophysical origins of the 
r-process and the associated Sr yields.

\subsection{Burst Scenarios}

Fig.~\ref{fig:bursts_srfe} shows [Sr/Fe] evolution and [Sr/Fe] distributions 
for our fiducial gas-driven and efficiency-driven starbursts, which can be 
compared to the [O/Fe] result in the top and bottom rows of 
Fig.~\ref{fig:fiducial_cases}. For ease of interpretation we have used the 
$y_\text{Sr}^\text{CC} = 3.5\times10^{-8}$ model for CCSN yields, and the 
black curves representing the unperturbed model are the same as the black 
curves in Fig.~\ref{fig:sr_yields} but shown with a zoomed-in axis range. 
In the gas-driven models, dilution with pristine gas first drives [Fe/H] lower 
at fixed [Sr/Fe]. For the burst at $t = 2$ Gyr, this backward jump is followed 
by a small upward hook, reminiscent of the behaviour of this model in [O/Fe]. 
However, this burst occurs very near the [Sr/Fe] ratio associated with the 
adopted CCSN yields, suggesting that CCSNe associated with the burst do not 
significantly modify the ISM [Sr/Fe]. Instead, it is likely that this increase 
is due to Sr production in AGB stars from earlier epochs. Subsequently, the 
detailed shape of the trajectory becomes complex as both SN Ia and AGB 
enrichment with metallicity dependent yields become important, and eventually 
it rejoins the trajectory of the unperturbed model. 
\par 
The $t = 5$ Gyr burst occurs after the maximum [Sr/Fe], produced because a 
$t^{-1.1}$ SN Ia DTD produces Fe on timescales longer than AGB stars produce 
Sr. In this model, [Sr/Fe] initially evolves downward following the addition 
of zero metallicity gas, both because of these late SNe Ia from previous 
generations of stars and because this is in the direction of the CCSN ratio of 
[Sr/Fe]$\approx$-0.1. Unfortunately, all of these excursions are small, and 
the impact on [Sr/Fe] distributions is almost negligible. Detecting the 
signature of these complex tracks would require correlating precise [Sr/Fe] 
and stellar age measurements. 
\par 
The impact of efficiency-driven bursts (lower panels) is somewhat stronger. 
Here the bursts drive upward excursions in [Sr/Fe] because both the CCSN and 
AGB channels contribute Sr faster than SN Ia Fe, and the slight boost of [Fe/H] 
increases the AGB yield. As seen previously in [O/Fe], the suppressed SFR after 
$\tau_*$ returns to its original value causes a downward hook in [Sr/Fe], as 
SN Ia Fe from stars produced during the burst dominates over the reduced CCSN 
and AGB contributions. These models produce larger deviations in the [Sr/Fe] 
distributions than the gas-driven models, with peaks at higher and lower 
[Sr/Fe] associated with the mid-burst maximum and post-burst minimum. However, 
the separation between these peaks is below 0.1-dex, so precise measurements 
would be needed to detect this signature. 

\begin{figure*} 
\includegraphics[scale = 0.32]{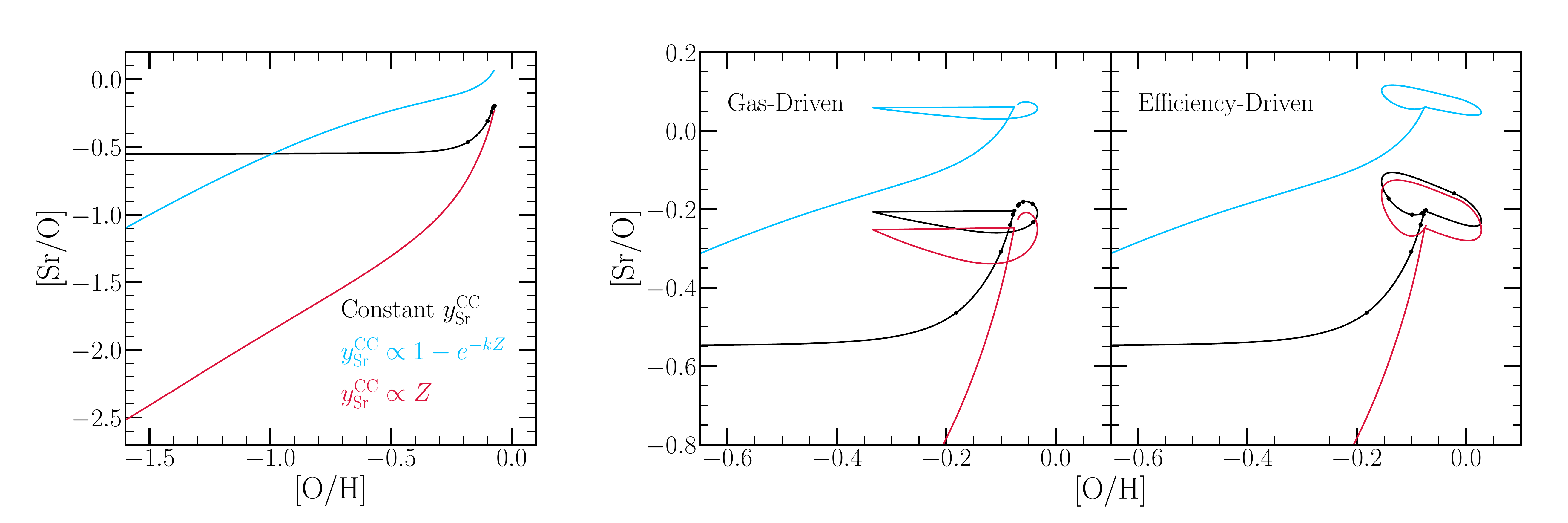} 
\caption{
Evolutionary tracks in the [Sr/O]-[O/H] plane for our fiducial burstless 
model (left) and for our fiducial gas-driven (middle) or efficiency-driven 
(right) starburst models, for three models of the $y_\text{Sr}^\text{CC}$ yield 
as labeled. In contrast to Figs.~\ref{fig:sr_yields} and~\ref{fig:bursts_srfe}, 
these results are independent of SN Ia enrichment, simplifying interpretation. 
In the middle panels, tracks initially evolve to lower [O/H] because of gas 
dilution, while in the right panel they evolve to higher [O/H] because 
increased SFE reduces the gas supply. The [Sr/O] evolution is driven mainly by 
the metallicity dependence of the Sr yields. In all panels, 
points are plotted at 1-Gyr intervals on models shown in black. 
}
\label{fig:sro_bursts} 
\end{figure*} 

The interpretation of Fig.~\ref{fig:bursts_srfe} is complicated partly by the 
fact that three enrichment processes are involved: CCSN, SN Ia, and AGB. 
Fig.~\ref{fig:sro_bursts} examines trajectories of [Sr/O] vs. [O/H], which are 
independent of SN Ia, at least given our assumption that SN Ia yields of O and 
Sr are insignificant. Here we show trajectories for our two 
metallicity-dependent CCSN yield models as well as the constant yield model. 
Tracks with smooth star formation (left panel) resemble the [Sr/Fe]-[Fe/H] 
tracks in Fig.~\ref{fig:sr_yields}, but without the dips coming for SN Ia Fe. 
For a gas-driven burst at $t = 5$ Gyr (middle panel), trajectories jump to 
lower [O/H] through dilution, then loop downward because the burst initially 
raises the rate of CCSN relative to AGB enrichment. These loops are analogous 
to the upward loops of [O/Fe], but O is now in the ratio denominator, and 
the timescales are CCSN vs. AGB rather than CCSN vs. SN Ia. The loop is flatter 
for the rotating star yield model because AGB stars make a smaller fractional 
contribution to Sr enrichment, and the CCSN contribution is boosted for both 
Sr and O during the burst. All trajectories eventually return to the late-time 
equilibrium of the unperturbed model. 
\par 
For an efficiency-driven burst at $t = 5$ Gyr (right panel), evolutionary 
tracks have a ``balloon-on-string'' appearance that can be understood as 
follows. By the time of the burst, the oxygen abundance has evolved to 
equilibrium, with 
\begin{equation} 
\label{eq:mdot_o_eq} 
\dot{M}_\text{O} \approx y_\text{O}^\text{CC}\dot{M}_* - 
(1 + \eta - r_\text{inst})\dot{M}_*(M_\text{O}/M_\text{ISM}) = 0 
\end{equation} 
where $M_\text{O}$ and $M_\text{ISM}$ are the oxygen and total mass in the ISM, 
respectively, and the oxygen abundance is 
\begin{equation} 
Z_\text{O,eq} = \left(\frac{M_\text{O}}{M_\text{ISM}}\right)_\text{eq} = 
\frac{y_\text{O}^\text{CC}}{1 + \eta - r_\text{inst}} 
\end{equation} 
\citepalias[][equations (11) and (14)]{Weinberg2017}. Boosting the star 
formation efficiency does not initially perturb $\dot{M}_\text{O}$ from 
zero because the sources and sinks are both proportional to $\dot{M}_*$, but 
the ISM gas mass decreases because of more rapid consumption, so 
$Z_\text{O} = M_\text{O}/M_\text{ISM}$ rises. The [Sr/O] ratio drops slightly 
at first because CCSN enrichment has increased relative to AGB enrichment, but 
the increased metallicity boosts the AGB Sr yield, so [Sr/O] loops upward once 
AGB enrichment from the starburst becomes important. As the burst evolves 
further, sinks exceed sources in equation~\refp{eq:mdot_o_eq}, so the [O/H] 
ratio evolves backward to lower values because $Z_\text{O} > Z_\text{O,eq}$. 
This evolution ``overshoots'' the original [O/H] equilibrium as the gas supply 
evolves back to its original value. Lower metallicity in turn leads to a drop 
in Sr yields and [Sr/O]. Eventually all models evolve back to the original 
pre-burst equilibrium. The loop of the $y_\text{Sr}^\text{CC} \propto Z$ model 
is widest in the [Sr/O] dimension because for this model AGB and CCSN yields 
both change with metallicity, and CCSN enrichment dominates over AGB enrichment 
in the $y_\text{Sr}^\text{CC} \propto 1 - e^{-kZ}$ model. 
\par 
Figs.~\ref{fig:bursts_srfe} and~\ref{fig:sro_bursts} show that the evolution of 
an AGB s-process element can be intricate because of both the intermediate 
timescale of AGB enrichment and metallicity dependent yields. Unfortunately the 
perturbations of [Sr/Fe] and [Sr/O] ratios are relatively small, so diagnosing 
starbursts with these ratios will require precise abundance measurements 
and reasonably precise stellar ages. 

\section{Long Term Modulation of Star Formation Rates}
\label{sec:oscillatory} 

\begin{figure*} 
\includegraphics[scale = 0.32]{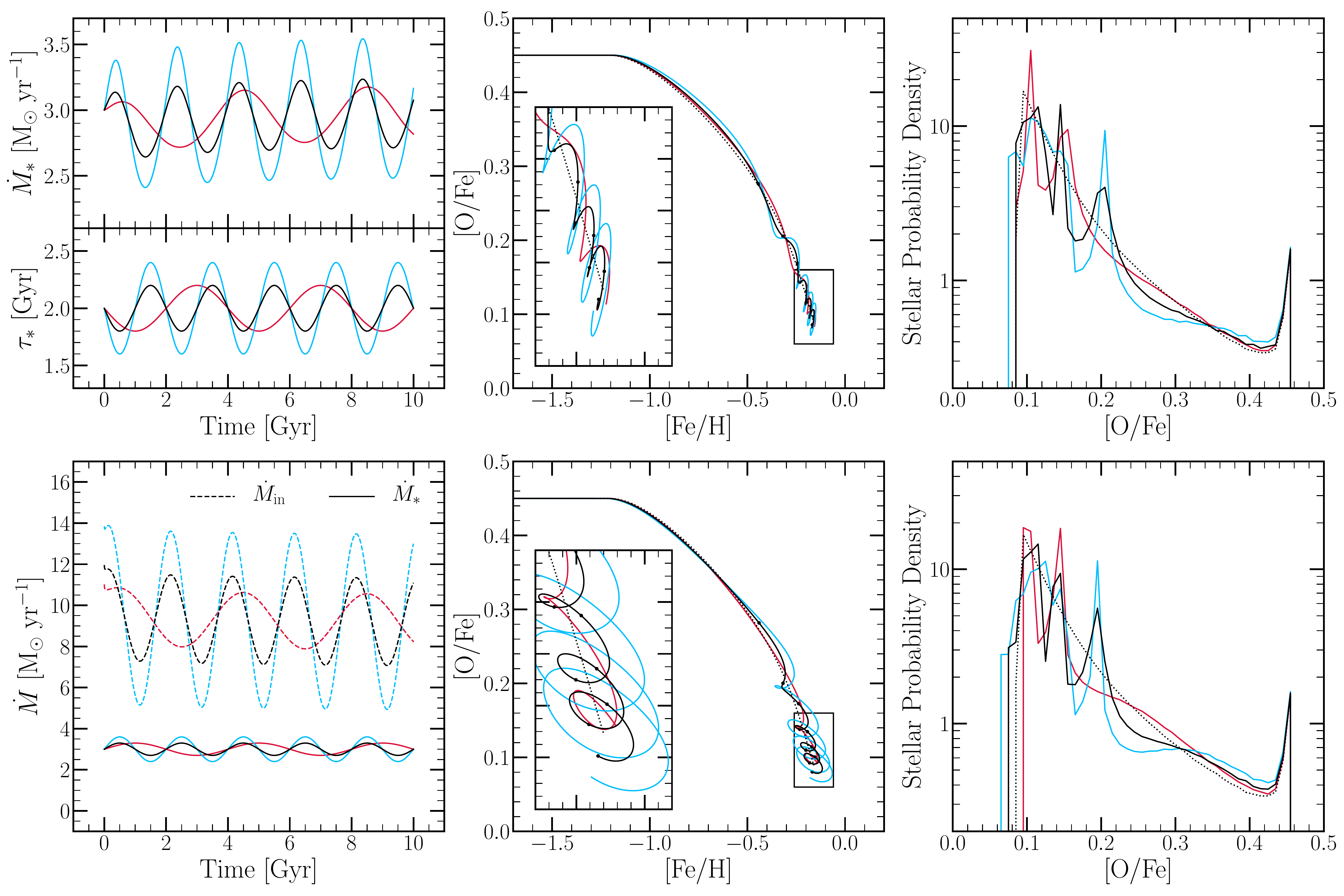} 
\caption{
Models with sinusoidal modulations of the SFR induced by modulations of the 
SFE timescale $\tau_*$ (top) or the gas infall rate $\dot{M}_\text{in}$ 
(bottom). Black curves represent a model with 10\% SFR modulations and a 2 Gyr 
period, while blue and red curves show the effect of doubling the amplitude or 
period of the modulation, respectively. In the middle and right panels, dotted 
black curves show results for our fiducial unperturbed model for comparison. 
In the middle panels, points are plotted at 1-Gyr intervals for the 10\% 
amplitude, 2-Gyr period model.
} 
\label{fig:oscil} 
\end{figure*} 

\begin{figure*} 
\includegraphics[scale = 0.32]{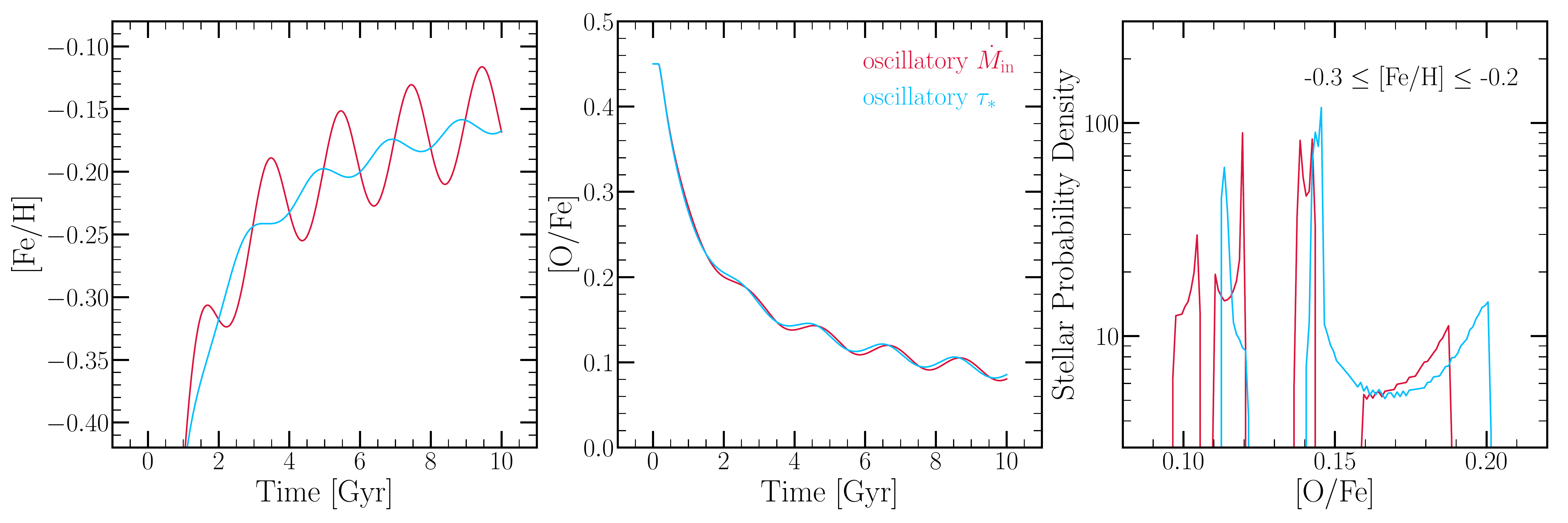} 
\caption{
Time evolution of [Fe/H] (left) and [O/Fe] (middle) for the 20\% amplitude, 2 
Gyr models driven by $\tau_*$ modulations (red curves) or $\dot{M}_\text{in}$ 
modulations (blue curves). Oscillations of [Fe/H] are larger in the infall 
modulation model, but oscillations of [O/Fe] are similar in the two models. 
The right hand panel shows the [O/Fe] distributions for stars in the range 
-0.2$\leq$[Fe/H]$\leq$-0.3, demonstrating that at constant [Fe/H], the 
resultant [O/Fe] distribution is not a simple bimodal gaussian. 
} 
\label{fig:oscil_v_time} 
\end{figure*}

\begin{figure*} 
\includegraphics[scale = 0.4]{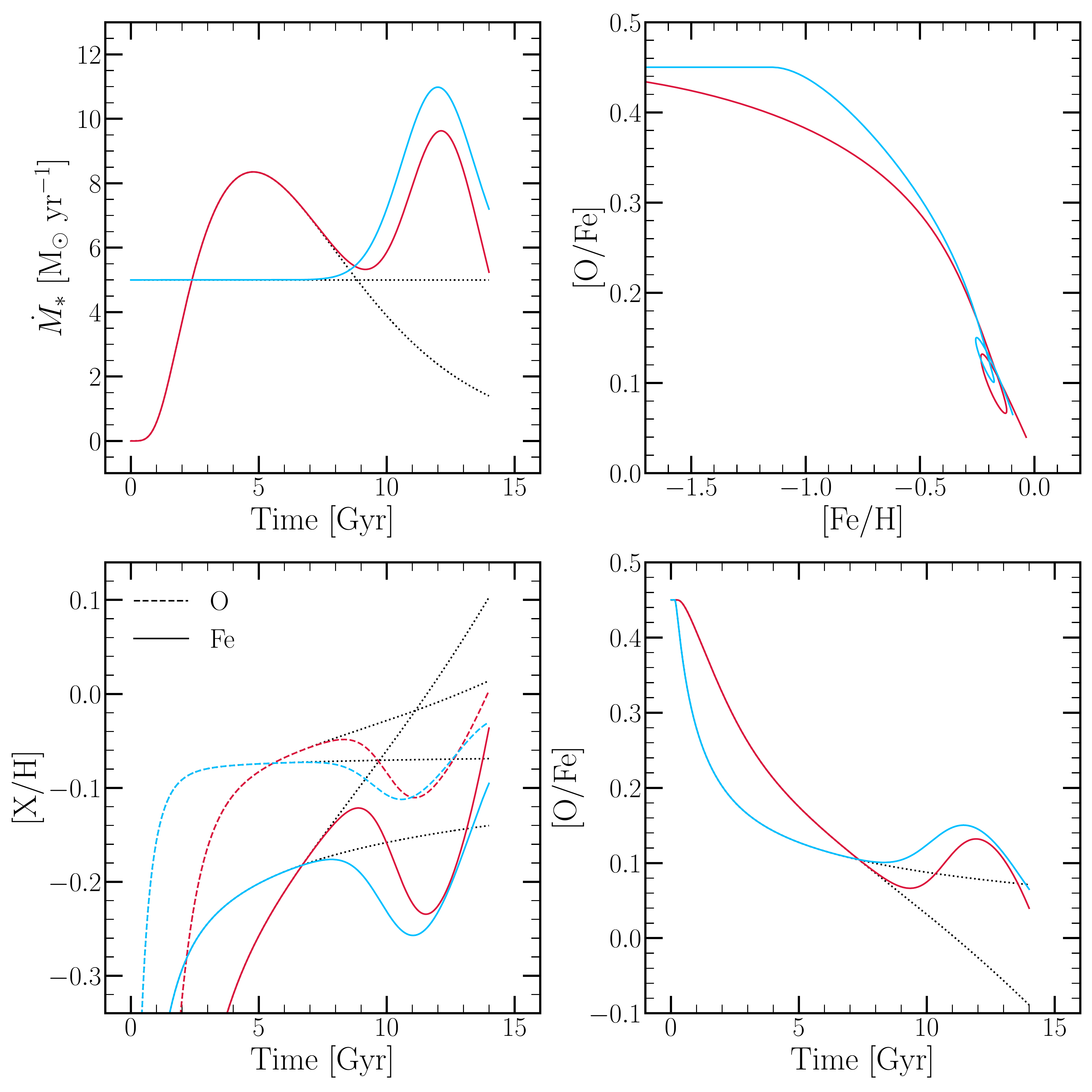} 
\caption{
Models loosely motivated by recent findings of a slow starburst in the Milky 
Way~$\sim$2 Gyr ago~\citep{Mor2019, Isern2019}. These models exhibit a 
constant SFR (blue) and a $\propto t^2e^{-t}$ infall history (red), to which 
we add a Gaussian centered at $t$ = 12 Gyr with dispersion $\sigma$ = 1 Gyr to 
both models, roughly doubling the SFR at its peak. Black dotted lines in all 
panels show the corresponding quiescent scenario, to which we add no starburst. 
\textbf{Top Left}: The SFR as a function of time. 
\textbf{Top Right}: The [O/Fe]-[Fe/H] tracks. We omit the tracks of the 
quiescent model from this panel for clarity. 
\textbf{Bottom Left}: [O/H] (dashed) and [Fe/H] (solid) as a function of time. 
\textbf{Bottom Right}: [O/Fe] as a function of time. 
These models produce mildly $\alpha$-enhanced stars with young ages and a low 
median age of stars near solar metallicity. 
} 
\label{fig:slow_bursts} 
\end{figure*} 

Bursts of star formation change the ratio of CCSNe to SNe Ia, producing loops 
in [O/Fe]-[Fe/H] trajectories and multiple peaks in [O/Fe] distributions. 
Slower, continuous variations of SFR also perturb the CCSN/SN Ia ratio in ways 
that can add complexity to these trajectories and distributions. From the 
standpoint of starbursts, such variations can be thought of as emulating a 
series of minor bursts throughout a galaxy's history, and are a possible 
source of scatter in [O/Fe] at fixed [Fe/H] in observed stellar populations. 
In the Milky Way disk,~\citet{BertranDeLis2016} estimate the intrinsic scatter 
in [O/Fe] as 0.03-0.04 dex in both the high-$\alpha$ (``chemical thick disk'') 
and low-$\alpha$ (``chemical thin disk'') stellar populations. 
\par
Fig.~\ref{fig:oscil} shows evolutionary tracks and [O/Fe] distributions for 
models with sinusoidal perturbations in SFR relative to a constant SFR model. 
In the upper panels, we create the SFR variations by modulating the SFE 
timescale $\tau_*$ about its baseline value of 2 Gyr. The black curve shows a 
model in which the amplitude of modulation is 10\% (i.e. 0.2 Gyr) and the 
period of modulation is 2 Gyr. Blue and red curves show models with a 20\% 
amplitude and a 4 Gyr period, respectively. The gas infall rate 
$\dot{M}_\text{in}$ and the outflow efficiency $\eta$ are held constant at 
their fiducial values. 
\par 
As one might expect from our efficiency-driven starburst models, each 
oscillation in $\tau_*$ induces a low amplitude loop in the [O/Fe]-[Fe/H] 
trajectory. For a 2 Gyr period, the first minimum in SFR occurs when [Fe/H] 
$\approx$ -0.45, and at lower metallicties the trjectory is only slightly 
different from that of an unperturbed model. At higher metallicities, there 
is a local maximum in [O/Fe] associated with each maximum in SFR, as one can 
see from the inset in the middle panel. The resulting [O/Fe] distributions 
have multiple peaks and troughs associated with flat and steep portions of the 
[O/Fe] trajectories, though these peaks can merge with each other into 
broader features. The peaks are sharper for higher amplitude modulations as 
expected. For the 10\% modulation, 2 Gyr period model there are three distinct 
peaks at [O/Fe] $\approx$ +0.11, +0.15, and +0.20, respectively, while the 
model with 4 Gyr period produces two distinct peaks at [O/Fe] $\approx$ +0.10 
and +0.15. 
\par 
The lower panels of Fig.~\ref{fig:oscil} show models in which we modulate 
the gas infall rate $\dot{M}_\text{in}$ while keeping $\tau_*$ and $\eta$ 
fixed. For these models we have chosen to modulate the \textit{star formation 
rate} $\dot{M}_*$ by 10\% or 20\% with a 2 or 4 Gyr period (solid curves in 
left panel). \texttt{VICE} automatically solves for the required modulations in 
$\dot{M}_\text{in}$ (dashed curves), which have the same period as the SFR 
modulations but a different phase and larger fractional amplitude. These gas 
supply modulations produce loops in [O/Fe] trajectories that resemble those of 
our infall-driven burst models. In particular, trajectories first move to lower 
[Fe/H] because of dilution, then to higher [O/Fe] and [Fe/H] because of 
subsequent star formation. The resulting [O/Fe] distributions show a multi-peak 
structure like that in the $\tau_*$-modulation models, with peaks at similar 
locations. 
\par 
Fig.~\ref{fig:oscil} shows that moderate amplitude fluctuations (10-20\%) of 
the SFR can produce a spread of [O/Fe] values at fixed [Fe/H], at 
the~$\sim$0.05-dex level. For $\tau_*$ modulations, this scatter appears 
mainly in the [O/Fe] dimension while for $\dot{M}_\text{in}$ modulations it 
appears in both the [O/Fe] and [Fe/H] dimensions, but the impact on the [O/Fe] 
distributions is similar. This is demonstrated further in the left and middle 
panels of Fig.~\ref{fig:oscil_v_time}, which compares the 20\% amplitude, 2-Gyr 
period models of the two modes of oscillations. The middle panel shows that 
both modes of oscillation produce strikingly similar evolution of the ISM 
[O/Fe] with time, but the oscillatory $\dot{M}_\text{in}$ model predicts much 
stronger oscillations in [Fe/H]. 
\par 
We also demonstrate in Fig.~\ref{fig:oscil_v_time} that these moderate 
variations do not produce a bimodal distribution in [O/Fe] at fixed [Fe/H] 
as observed in the Milky Way; a more dramatic departure from this class of 
models is required. The right panel shows the normalized stellar MDFs in [O/Fe] 
only considering stars with -0.3 $\leq$ [Fe/H] $\leq$ -0.2, and although these 
display complex structure, neither model reproduces the [O/Fe] distribution 
found by, e.g.,~\citet{BertranDeLis2016}, which is well described by two 
Gaussians separated by~$\sim$0.15 dex. It is also notable that these SFR 
modulations only induce scatter in [O/Fe] at locations well beyond the knee 
of the [O/Fe]-[Fe/H] track. In part this is because our chosen parameters 
predict models which evolve past the knee quickly, before a full cycle of the 
SFR modulation. However, enrichment near the [O/Fe] plateau is dominated by 
CCSN in any case, so fluctuations in the SFR that change the CCSN rate have 
little leverage on [O/Fe]. Explaining intrinsic scatter in [O/Fe] (or ratios 
for other $\alpha$-elements) near the plateau of the high-$\alpha$ sequence 
requires a different mechanism, such as incomplete mixing of CCSN ejecta that 
individually have varying [O/Fe] ratios. 

\section{Slow Starburst in the Milky Way} 
\label{sec:slowburst} 

Although our starburst models are most obviously relevant to dwarf galaxies 
with episodic star formation histories, some recent observations suggest that 
the Milky Way itself experienced substantially elevated star formation 2-3 Gyr 
ago. \citet{Mor2019} infer such a history by comparing population synthesis 
models to observed stellar luminosity functions and color-magnitude diagrams 
from Gaia data. \citet{Isern2019} reaches similar conclusions from modeling the 
luminosity function of white dwarfs in the solar neighborhood measured using 
Gaia parallaxes. Although these white dwarfs are close to the sun, dynamical 
mixing implies that they at least sample the history of the solar annulus, 
and older white dwarfs likely sample a range of galactocentric radii because of 
radial mixing. Resolved stellar population studies of the M31 disk also provide 
evidence for elevated star formation 2-4 Gyr ago with much lower SFR before 
and after~\citep[][Figs. 22-23]{Williams2017}. 
\par 
Fig.~\ref{fig:slow_bursts} presents the evolution of two models loosely 
motivated by these observations. The first (blue curves) has a constant SFR to 
which we have added a burst described by a Gaussian centered at $t = 12$ Gyr 
(lookback time 2 Gyr) with dispersion of 1 Gyr. At its peak, this burst 
approximately doubles the galaxy SFR relative to the pre-burst value. The 
second model (red curve) adds a similar burst to a model with an infall 
history described by $\dot{M}_\text{in} \propto t^2e^{-t}$ with $e$-folding 
timescale $\tau_\text{inf} = 2.2$ Gyr. The pre-burst SFR first climbs as the 
gas supply builds (starting from zero), then declines as the infall rate slows. 
The qualitative appearance of this model is similar to those in Fig. (1) 
of~\citet{Isern2019}. For both models we adopt $\eta$ = 2.5 and 
$\tau_* = (\text{2 Gyr})(M_\text{g}/6.0\times10^9\ M_\odot)^{-0.5}$, allowing 
\texttt{VICE} to solve for the infall rate required to produce the starburst. 
We also plot the evolution of the corresponding quiescent models in dotted 
lines for comparison; they follow the same evolution but with no added 
starburst. 
\par 
The [O/Fe]-[Fe/H] tracks show loops similar to those of our infall-driven and 
efficiency-driven burst models (e.g., Fig.~\ref{fig:fiducial_cases}). As found 
in previous studies~\citepalias{Andrews2017,Weinberg2017}, the exponential 
infall model exhibits slower pre-burst evolution and a more gradual ``knee'', 
and because of the short $e$-folding timescale even the unperturbed model does 
not approach equilibrium by $t$ = 14 Gyr. The critical feature of these models 
relative to our fiducial strabursts is that the effects of the bursts have not 
decayed by the end of the simulations at $t$ = 14 Gyr. Over the final 2 Gyr, 
the values of [O/H] and [Fe/H] are rapidly climbing and end at values higher 
than those reached at any previous time. The [O/Fe] values in both models 
reach a local maximum at $t$ = 12 Gyr, then fall for the final 2 Gyr. 
\par 
These results are of particular interest in light of recent 
studies of age-abundance relations from the Apache Point Observatory Galaxy 
Evolution Experiment (APOGEE)~\citep[e.g.][]{Martig2016,
SilvaAguirre2018, Feuillet2018, Feuillet2019}. The late bump in [O/Fe] could 
help explain populations of young $\alpha$-enhanced stars~\citep{Martig2016, 
Feuillet2019}, though in these models such stars would have modest 
$\alpha$-enhancements, 
near-solar metallicity, and age $\approx$ 2 Gyr. The late-time bumps in [O/H] 
and [Fe/H] could help to explain the strikingly young (1-2 Gyr) median ages that 
\citet{Feuillet2018} find for solar neighborhood stars with [Fe/H] $\approx$ 
0 or [O/H] $\approx$ 0. Finally, the most $\alpha$-poor stars predicted by 
these models form at late times in the wake of the burst, potentially 
explaining the low median age ($\sim$1 Gyr) that~\citet{Feuillet2018} find for 
stars with [$\alpha$/Fe] < 0. The age-metallicity relation 
for solar neighborhood stars exhibits large scatter~\citep{Edvardsson1993}, 
and explaining this scatter likely requires radial mixing of stellar 
populations~\citep[e.g.][]{Schoenrich2009} or some other mechanism not 
represented in one-zone GCE models. However, while multi-zone models with 
radial mixing and smooth star formation histories can explain a large 
dispersion in age-abundance relations, they still have difficulty reproducing 
the young median ages inferred for solar metallicity stars~\citep[][see 
their Fig. 15]{Feuillet2018}. 
The one-zone models presented here suggest that elevated star formation in the 
recent past could have a significant impact on age-abundance relations, 
pushing them away from the equilibrium behaviour predicted for smooth star 
formation histories. Furthermore, the differences between these models and 
their corresponding quiescent cases at late times raise the intriguing 
possibility that the recent burst in the Milky Way has not yet fully decayed. 
This would imply that the present day chemistry of the Milky Way is still 
mildly perturbed due to the recent starburst. We reserve an exploration of 
models combining radial mixing with star formation histories like those 
of~\citet{Mor2019} and~\citet{Isern2019} for future work. 
\par 

\section{Conclusion}
\label{sec:conclusion} 

We have studied one-zone chemical evolution models tracking the enrichment of 
oxygen, iron, and strontium with the goal of understanding the impact of star 
formation bursts on elemental abundance ratios. To this end, we have developed 
the \texttt{Versatile Integrator for Chemical Evolution} (\texttt{VICE}), a 
\texttt{python} package optimized for handling highly non-linear chemical 
evolution models. With this new tool, we first simulated gas-driven 
starbursts, whereby an amount of gas comparable to the current ISM mass of a 
galaxy is added to the ISM on timescales shorter than the depletion time. 
These starburst models predict hooks in the [O/Fe]-[Fe/H] plane; the rapid 
addition of pristine gas first causes a reduction in [Fe/H] at fixed [O/Fe], 
then the elevated rate of CCSNe relative to SNe Ia drives the ISM to higher 
[O/Fe] and [Fe/H], and finally the onset of SNe Ia associated with the 
starburst pushes the ISM back toward the [O/Fe]-[Fe/H] track of the unperturbed 
(i.e. no-burst) model. The rate at which extra gas is added to the galaxy 
affects the detailed shape of these jump-and-hook trajectories. 
\par 
Our unperturbed constant-SFR models predict one peak in the [O/Fe] distribution 
at the ``plateau'' ratio of CCSN O and Fe yields, and a second peak associated 
with the late-time equilibrium in which CCSN and SN Ia rates are equal. Our 
gas-driven starburst models predict a third peak in the [O/Fe] distribution, 
associated with stars that form out of the $\alpha$-enhanced ISM 
during/following the burst before SNe Ia have driven evolution back toward the 
unperturbed evolutionary track. The peak is centred near the value of [O/Fe] 
at the top of the hook in the [O/Fe]-[Fe/H] trajectory, and its location and 
shape are insensitive to the timescale on which the gas is added, provided 
that this timescale is short compared to the depletion time. Earlier 
starbursts produce this third peak at higher [O/Fe] because they arise when 
the starting value of [O/Fe] is further from its eventual equilibrium. Thus, 
even without accurate ages for individual stars, the existence of extra 
peaks in the [O/Fe] distribution (or [X/Fe] distribution for other 
$\alpha$-elements) can provide an observable diagnostic for past bursts of 
galactic star formation, and the locations of these peaks can provide 
estimates of the timing of these bursts. 
\par 
A gas-driven starburst could arise from the merger of a gas rich system or a 
temporary increase in accretion rate. A starburst can also arise from a 
temporary increase in star formation efficiency, consuming the available gas 
more quickly, perhaps because of a dynamical disturbance that does not 
increase the gas supply. The evolutionary tracks of efficiency-driven 
starbursts differ in form from those of gas-driven starbursts, first because 
there is no drop in [Fe/H] before the increase in [O/Fe], and second because 
[O/Fe] loops below the track of the unperturbed model once the post-burst 
gas supply is depleted, which allows the CCSN rate to fall well below the rate 
of SNe Ia from stars that formed during the burst. With sufficiently precise 
data, the [O/Fe] distribution of an efficiency-driven burst can be 
distinguished from that of a gas-driven burst, in part by the shape of the 
[O/Fe] peak for $\alpha$-enhanced stars formed during the burst, and in part 
by the presence of an additional population of $\alpha$-deficient stars. 
\par 
In short, the chemical response to a simple starburst is driven by a 
perturbation of the CCSN and SN Ia rates. Conventionally, one-zone models tie 
the outflowing wind to the instantaneous star formation rate (i.e. 
$\dot{M}_\text{out} = \eta\dot{M}_*$), which in turn ties it to the CCSN rate. 
If SNe Ia contribute to the outflowing wind, then a better approximation of 
the outflow rate would be one that is tied to a time-averaged SFR (i.e. 
$\dot{M}_\text{out} = \eta\langle\dot{M}_*\rangle_{\tau_\text{s}}$), where 
$\tau_\text{s}$ is the outflow smoothing time. Nonzero $\tau_\text{s}$ allows 
the ISM to retain more gas at the onset of a starburst, because the outflow is 
more sensitive to the preburst SFR. The ISM is then gas-poor in the decay of 
the starburst, because the outflow is most sensitive to the elevated SFR from 
the recent burst. Varying $\tau_\text{s}$ between 0 and 1 Gyr has minimal 
impact on the predicted [O/Fe]-[Fe/H] trajectory or the stellar [O/Fe] 
distribution of gas-driven starbursts. However, efficiency-driven burst models 
with $\tau_\text{s}$ = 0.5 - 1 Gyr exhibit wider loops and produce lower 
[$\alpha$/Fe] stars than in the $\tau_\text{s}$ = 0 model. These 
$\alpha$-deficient stars are produced late in the burst when the ISM is 
gas-poor, an effect that is magnified by a nonzero $\tau_\text{s}$ because 
the gas outflow rate is higher and the CCSN rate is lower. 
\par 
While our simplest starburst scenarios are either gas- or efficiency-driven, 
there is observational evidence for starbursts driven by both 
an increase in the gas supply and an increase in the 
efficiency~\citep[][and the citations therein]{Kennicutt2012}. As a simple 
example of a ``hybrid'' starburst, we considered a model with a rapid influx 
of gas and an SFE timescale $\tau_* \propto M_\text{g}^{-1/2}$ as suggested 
by the Kennicutt-Schmidt law~\citep{Schmidt1959, Schmidt1963, Kennicutt1998}. 
For $\tau_\text{s}$ = 0 or 0.5 Gyr, the [O/Fe]-[Fe/H] tracks of this model are 
nearly the same as those of the gas-driven constant-$\tau_*$ model. For 
$\tau_\text{s}$ = 1 Gyr, the hybrid model shows aspects of both gas-driven and 
efficiency-driven models, including a population of $\alpha$-deficient stars. 
\par 
The AGB models of~\citet{Cristallo2011} predict Sr yields that are strongly 
dependent on metallicity and dominated by 2 - 4 $M_\odot$ stars. Predicted 
CCSN yields of Sr are sensitive to rotationally induced mixing; 
the~\citet{Limongi2018} yields for non-rotating progenitors vs. progenitors 
with $v_\text{rot}$ = 150 km s$^{-1}$ differ by 1-2 orders of magnitude, with 
strong but differing metallicity dependence. Near solar metallicity, the AGB 
yields and non-rotating CCSN yields are comparably important, but the 
$v_\text{rot}$ = 150 km s$^{-1}$ CCSN yields would outweight AGB yields by a 
large factor. 
\par 
Reproducing the approximately flat trend of [Sr/Fe] vs. [Fe/H] found in the 
Milky Way and in dwarf satellites~\citep{Mishenina2019, Hirai2019} requires 
an additional source of Sr with a yield that is nearly independent of 
metallicity, perhaps the neutrino-driven winds from newly formed neutron 
stars~\citep{Thompson2001, Vlasov2017, Thompson2018}. For any of these CCSN 
yield models, the tracks of [Sr/Fe] vs. [Fe/H] in starburst models are complex, 
affected by the yield metallicity dependence and by the differing timescales 
of CCSN, AGB, and SN Ia enrichment. Tracks of [Sr/O] vs. [O/H] are simpler 
because they are independent of SN Ia enrichment. However, the total range 
of [Sr/Fe] or [Sr/O] induced by starbursts is small, typically 0.05-0.1 dex, 
and in combination with yield uncertainties this small dynamic range makes it 
difficult to use Sr abundances as a diagnostic of starburst behavior. 
\par 
In addition to strong (factor of~$\sim$2) starbursts, we have investigated 
models with 10-20\% sinusoidal modulations of a constant SFR, induced by 
variations in infall rate or star-formation efficiency. These models predict 
[O/Fe]-[Fe/H] tracks that oscillate about the prediction of the constant SFR 
model. The produce a multi-peaked structure in [O/Fe] distributions, though 
in the presence of observational errors these peaks would likely merge into 
a broader distribution. These variations do not produce a bimodal [O/Fe] 
distribution, so they are not the origin of the observed separation of thin 
and thick disk sequences~\citep[e.g.][]{Bensby2003, Hayden2015, 
BertranDeLis2016}. However, moderate variations in SFR could be a source of 
scatter in [O/Fe] along these sequences. With our adopted parameter values, 
our smooth evolution models approximately reproduce the observed high-$\alpha$ 
sequence. SFR oscillations produce a spread of~$\sim$0.05-0.1 dex in [O/Fe] 
for [Fe/H] $\gtrsim$ -0.4, but they cannot produce scatter near the 
high-$\alpha$ plateau of this sequence because the enrichment of those stars 
is dominated by CCSN in any case. 
\par 
Motivated by findings on the recent star formation history of the 
Milky Way by~\citet{Mor2019} and~\citet{Isern2019}, we explored 
models that exhibit slow, factor of~$\sim$2 increases in the SFR at lookback 
times of~$\sim$2 Gyr, adopting a simple gaussian with dispersion of $\sigma$ = 
1 Gyr to describe the starburst. A late-time, slow starburst may help to 
explain otherwise puzzling features of the age-abundance relations observed 
in APOGEE~\citep{Martig2016,SilvaAguirre2018,Feuillet2018,Feuillet2019}, such 
as young stars with mild $\alpha$-enhancements and young median ages of solar 
metallicity or $\alpha$-deficient stars. Complete modeling of these 
observables requires multizone models that account for radial mixing of 
stellar populations, and we reserve such investigations to future work. 
\par 
Throughout this paper we have adopted an O yield similar to those predicted 
by~\citet{Chieffi2004} and~\citet{Chieffi2013}, assuming a Kroupa IMF in which 
all stars with $M > 8\ M_\odot$ explode. With this yield, evolving to solar 
metallicity requires fairly strong outflows, with $\eta$ = 2.5 
(e.g.,~\citealp{Finlator2008};~\citealp{Peeples2011};~\citetalias{Andrews2017}; 
\citetalias{Weinberg2017}). With lower IMF-averaged SN yields, which could 
arise if many massive stars form black holes instead of exploding, lower 
values of $\eta$ would be needed to reach the same final metallicity. Although 
results for lower yield, lower $\eta$ models would differ in detail from those 
presented here, mainly because the depletion time $\tau_\text{dep} = 
\tau_*/(1 + \eta - r_\text{inst})$ would be longer for the same $\tau_*$. We 
have investigated several of our models in which both yields and $\eta$ are 
reduced by a factor of~$\sim$2, and found that our qualitative conclusions 
still hold. 
\par 
We have released \texttt{VICE} as open-source software under the MIT 
licence. Source code, installation instructions, and documentation can be 
found at~\url{http://github.com/giganano/VICE.git}. We also include code that 
runs the simulations of our models and produces the figures in this paper. 

\section{Acknowledgements}
We thank Jenna Freudenburg, Jennifer Johnson, Tuguldur Sukhbold, Todd 
Thompson, and Fiorenzo Vincenzo for helpful advice. This work was supported 
by NSF grants AST-1211853 and AST-1909841 and by an Ohio State University 
graduate fellowship. 

\bibliographystyle{mnras}
\bibliography{draft2}
\label{lastpage}
\end{document}